\documentclass[a4paper,11pt,final]{article}

\usepackage{authblk}
\usepackage{graphics}
\usepackage{graphicx}
\usepackage{epsfig}
\usepackage{amssymb}
\usepackage{amsthm}

\usepackage{amsmath}
\date{}

\begin{document}

\title{Relativistic analysis of  stochastic kinematics}

\author{Massimiliano Giona}
\affil{Dipartimento di Ingegneria Chimica, Materiali, Ambiente\\
La Sapienza Universit\`a di Roma\\ Via Eudossiana 18, 00184 Roma, Italy\\
 E-mail: massimiliano.giona@uniroma1.it}
\maketitle

\begin{abstract}
The relativistic analysis
of stochastic kinematics is developed in order to
determine the transformation of the effective
diffusivity tensor in inertial
frames.  
Poisson-Kac  stochastic processes are initially considered.
 For   one-dimensional   spatial models,  
 the effective diffusion coefficient $D$ measured in a 
frame  $\Sigma$ moving with velocity $w$ with respect to the
rest frame of the stochastic
process can be expressed as $D= D_0 \, \gamma^{-3}(w)$.
Subsequently, higher dimensional processes are analyzed,
and it is shown that the
diffusivity tensor in a moving frame becomes non-isotropic
with $D_\parallel = D_0 \,  \gamma^{-3}(w)$, and $D_\perp = 
D_0 \, \gamma^{-1}(w)$,
where $D_\parallel$ and $D_\perp$ are the diffusivities parallel and
orthogonal to the velocity of the moving frame.
The analysis of discrete Space-Time Diffusion processes
permits to obtain a general transformation theory of the
tensor diffusivity, confirmed by several
different simulation experiments. Several implications of the
theory are also addressed and discussed.
\end{abstract}

\section{Introduction}
\label{sec1}

Merging stochastic dynamics within the formal
structure of relativity (special or general) is a relevant
issue in theoretical physics (field theory) with important
implication in high-energy physics and cosmology \cite{gen1,gen2}.
It is a challenging issue, due to causality and  to the formal constraints
of the Minkowskian structure of the space-time, imposed
by the boundedness of the propagation velocity of physical processes,
that forces to consider space and time variable on equal footing.

The limit imposed by the constant value of the  velocity  of light
{\em in vacuo}, implies the any consistent relativistic stochastic
process should possess bounded propagation velocity
and, as a consequence of this, an almost everywhere
smooth structure of the space-time trajectories.

The works by Dudley \cite{dudley1,dudley2,dudley3} 
and Hakim \cite{hakim1,hakim2,hakim3} elucidated further
the subtleties of the relativistic formulation of
stochastic processes. The most relevant constraint is the
impossibility of a strictly Markovian process in the 
Minkowski space-time. As observed by Dudley \cite{dudley2},
the structure of the Minkowsky space-time forces to include
information on the velocity in specifying the state of
a stochastic process.
This general and significant observation can be interpreted
and followed in two conceptually different ways in order to
define relativistic models of Brownian motion.

The first strategy is to build up a relativistic
version of Ornstein-Uhlenbeck processes in the $\mu$-space \cite{hakim1},
that is the Cartesian product of the Minkowski space-time
${\mathcal M}_4$ times the ${\mathbb R}^4$ space of the
4-velocities.
This is the strategy adopted in defining the
so-called Relativistic Ornstein-Uhlenbeck
process by Debbasch et al. \cite{debbasch1,debbasch2},
and the Relativistic Brownian Motion by Dunkel and H\"anggi \cite{dunkel1,dunkel2}. These classes of models represent Ornstein-Uhlenbeck processes,
i.e., Langevin equations for the position and the momentum of
a particle driven by a stochastic force in the
presence of a  friction contribution. The stochastic force is modelled
in the form of a Wiener process, and the momentum-dependent
prefactors accounting for the Lorentz covariance, modulating
both the friction and the intensity of the stochastic
perturbations,  can be  derived from
the stationary relativistic velocity distribution, expressed
by the Juttner distribution \cite{juttner} (or by
the modified Juttner distribution), corresponding to
the stationary solution of the relativistic Boltzmann equation
\cite{hakim_libro,cercignani,libby}. A central limit theorem for
these families of relativistic processes is developed in
\cite{angst}.

All these models describe a stochastic dynamics but not a stochastic
kinematics, i.e., a stochastic process involving exclusively 
space-time coordinates that, according to the
above observation by Dudley and Hakim cannot be grounded
on a strict Markovian model.
The way for approaching a relativistically consistent stochastic
kinematic model, is to consider a system of bounded velocities,
the selection of which is controlled by a Markov-chain process.
This is the essence of Poisson-Kac processes,
introduced by Marc Kac \cite{kac} and further elaborated
by many authors \cite{dicho1,dicho2,dicho3,giona_kac1}.
For a general review of the different approaches in
relativistic stochastic analysis see \cite{hanggi_rev}.

The aim of this article is to analyze the (special) relativistic
transformation of stochastic kinematics, and specifically the
transformation of the tensor diffusivity induced by a Lorentz
boost. Albeit this issue is of general relevance both
in theory and applications, it has  never been addressed in
the literature, to the best of the author's knowledge.
A possible explanation of this lack stems from
the fact that for relativistic stochastic
processes a $\mu$-space formulation has been
followed, and the resulting nonlinear Langevin equations
in ${\mathcal M}_4 \times {\mathbb R}^4$ are fairly complex
and not easily amenable to a closed-form analysis
of the associated diffusivity tensor.

In this article, using different approaches and relativistic  stochastic
processes  (Poisson-Kac processes, discrete Space-Time  Diffusion
models recently studied in \cite{giona_std1,giona_std2}, etc.)
the relativistic transformation of the tensor diffusivity
is derived and some of its implications explored.

The article is organized as follows. Section \ref{sec2} 
reviews briefly the class of Poisson-Kac processes,
and Section \ref{sec3} their relativistic transformation.
Section \ref{sec4} develops the moment analysis of the  spatial
one-dimensional
Poisson-Kac process in order to obtain the expression
for the effective diffusivity measured in two
inertial frames in relative motion. Section \ref{sec5}
extends the analysis to higher dimensions, showing
the occurrence of two different scalings for the
longitudinal and transversal diffusivities. Scaling
analysis and numerical simulations are developed
in Sections \ref{sec6} and \ref{sec8}, respectively.
Section \ref{sec8_bis} discusses a by-product of the
transformation analysis of tensor diffusivity, associated
with the relativistic invariance of the stochastic
action. Section \ref{sec9}
addresses the general transformation theory of tensor
diffusivity, by considering discrete  Space-Time Diffusion
models as a prototypical example of stochastic
processes amenable to closed-form analysis
due to the homogenization theory developed in \cite{giona_std1}.
Finally, Section \ref{sec10} discusses some implications
of the theory, including also a brief analysis
of the concept of deterministic vs stochastic motion
in a Minkowskian space-time.

\section{Poisson-Kac processes}
\label{sec2}

Let $\Sigma^\prime$, (space-time coordinates $x^\prime$, $t^\prime$) be
a frame in which a Poisson-Kac process is ``at rest'', i.e., it
possesses vanishing effective (long-term) velocity.
$\Sigma^\prime$ can be referred to as the {\em rest frame} of the
process.
In $\Sigma^\prime$ the free Poisson-Kac process is described by the
stochastic equation
\begin{equation}
d x^\prime(t^\prime) = b^\prime_0 (-1)^{\chi(t^\prime)} \, d t^\prime 
\label{eq1}
\end{equation}
where $\chi(t^\prime)$ is the realization of a Poisson process
possessing transition rate $a_0$, i.e., such that
the probability density function $p_{\tau^\prime}(\tau^\prime)$
for the switching times of the Poisson process is given
by the exponential distribution $p_{\tau^\prime}(\tau^\prime)=
a_0^\prime \, e^{-a_0^\prime \, \tau^\prime}$, $\tau^\prime \in [0,\infty)$.

The parameter $b^\prime_0$ is the characteristic local
velocity of the Poisson-Kac perturbation, $0 < b^\prime \leq c$,
where $c$ is the light velocity {\em in vacuo}.

Let  $p^\prime(x^\prime,t^\prime)$ be the probability density function
associated with the stochastic evolution (\ref{eq1}),
and $p^{\prime,\pm}(x^\prime,t^\prime)$ the partial
probabilities characterizing the statistical
evolution of the Poisson-Kac process (also referred to as {\em partial
probability waves}).
In $\Sigma^\prime$, the equations for $p^{\prime, \pm}(x^\prime,t^\prime)$
read
\begin{eqnarray}
\frac{\partial p^{\prime,+}}{\partial t^\prime} & = & - b^\prime_0
\, \frac{\partial p^{\prime,+}}{\partial x^\prime} - a_0^\prime \,  p^{\prime,+}
+ a_0^\prime \,  p^{\prime,-} \nonumber \\
\frac{\partial p^{\prime,-}}{\partial t^\prime} & = &  b^\prime_0
\, \frac{\partial p^{\prime,-}}{\partial x^\prime} + a_0^\prime \,  p^{\prime,+}
-  a_0^\prime \, p^{\prime,-} 
\label{eq2}
\end{eqnarray}
where $p^\prime=p^{\prime,+}+p^{\prime,-}$.
The parameter $D_0$ given by
\begin{equation}
D_0 = \frac{(b^\prime_0)^2}{ 2 \, a_0^\prime}  
\label{eq3}
\end{equation}
represents the diffusion coefficient of the Poisson-Kac process
in the rest frame ({\em rest diffusion coefficient}).
It is well known from the work by 
 Kac  \cite{kac} that in the limit of $b^\prime_0$ and $a_0^\prime$ 
tending to infinity,
keeping fixed the ratio $D_0$, the solution of eq. (\ref{eq2})   converges
in the long-term limit to that of a  pure
diffusion equation for $p^\prime(x^\prime,t^\prime)$ 
characterized by the diffusivity $D_0$.

\section{Inertial transformations}
\label{sec3}

Let $\Sigma$ be an inertial frame moving with  constant velocity $w$, $|w| \leq c$ with
respect to $\Sigma^\prime$ and $(x,t)$ its space-time coordinates.
Enforcing the Lorentz transformation between $\Sigma$ and $\Sigma^\prime$,
\begin{equation}
x= \gamma(w) \left ( x^\prime -w \, t^\prime \right ) \; ,
\qquad t= \gamma(w)  \left ( t - w \, x^\prime/c^2 \right ) 
\label{eq4}
\end{equation}
where $\gamma(w)=(1-w^2/c^2)^{-1/2}$ is the Lorentz factor, eqs. (\ref{eq1})
become
\begin{eqnarray}
\gamma(w) \,  \left ( 1- \frac{b^\prime_0  \, w}{c^2} \right )
\, \frac{\partial p^{\prime,+}}{\partial t}
& = & - \gamma(w) \, (b^\prime_0 - w)
\, \frac{\partial p^{\prime,+}}{\partial x}
- a_0^\prime \, p^{\prime,+} + a_0^\prime \, p^{\prime,-} \nonumber \\
\gamma(w) \,  \left ( 1 + \frac{b^\prime_0  \, w}{c^2} \right )
\, \frac{\partial p^{\prime,-}}{\partial t}
& = &  \gamma(w) \,  (b^\prime_0 +w)
\, \frac{\partial p^{\prime,-}}{\partial x}
+ a_0^\prime \, p^{\prime,+} - a_0^\prime \, p^{\prime,-}  
\label{eq5}
\end{eqnarray}
Let us define the transformed partial probability densities
$p^+(x,t)$, $p^-(x,t)$ in $\Sigma$ as
\begin{equation}
p^+ =    \gamma(w) \, 
\left ( 1- \frac{b^\prime_0  \, w}{c^2} \right ) \, p^{\prime,+} \; ,
\qquad
p^- =  \gamma(w) \, \left  ( 1+ \frac{b^\prime_0  \, w}{c^2} \right ) \, p^{\prime,-} 
\label{eq6}
\end{equation}
The quantities  $p^+(x,t)$ and $p^-(x,t)$ are the
representation of the partial probability densities for
the stochastic process eq. (\ref{eq1}) in $\Sigma$, and their
balance equations read as
\begin{eqnarray}
\frac{\partial p^+}{\partial t} & = &
 - b^+ \, \frac{\partial p^+}{\partial x}
- a^+ \, p^+ + a^- \, p^- \nonumber \\
\frac{\partial p^-}{\partial t} & = & 
  b^- \, \frac{\partial p^+}{\partial x}
+ a^+ \, p^+ - a^- \, p^-  
\label{eq7}
\end{eqnarray}
where
\begin{equation}
b^+ = \frac{b^\prime_0-w}{1- b^\prime_0 \, w/c^2} \; , \qquad
b^-= \frac{b^\prime_0+w}{1+ b^\prime_0 \, w/c^2} 
\label{eq8}
\end{equation}
and
\begin{equation}
a^+ = \frac{a_0^\prime}{\gamma(w) \, (1- b^\prime_0 \, w/c^2)} \; , \qquad
a^- = \frac{a_0^\prime}{\gamma(w) \, (1+ b^\prime_0 \, w/c^2)} 
\label{eq9}
\end{equation}
The physical interpretation of  eq. (\ref{eq6})
based on the covariance of the Poisson-Kac process  
is given in \cite{giona0}.

From eq. (\ref{eq7}), the overall probability density $p=p^++p^-$
in $\Sigma$ is a conserved quantity. Moreover, from the definition
(\ref{eq6}) it follows that the transformation for the
probability densities/fluxes in the two systems is 
given by
\begin{equation}
p  =  \gamma(w) \, \left ( p^\prime - \frac{w}{c^2} \, J^\prime
  \right ) \,  \qquad
J  =  \gamma(w) \, \left ( J^\prime - w p^\prime  \right )
\label {eq10}
\end{equation}
where $J^\prime= b_0^\prime \, (p^{\prime +}-p^{\prime -})$,
$J= b^+ p^+- b^- p^-$ are the probability fluxes in the
two reference systems. Eq. (\ref{eq10})
corresponds to the Lorentz boost for the two-dimensional
(as we consider a $1+1$ Minkowski space-time)
4-vector $j_\nu^\prime=(c \, p^\prime,J^\prime)$.
Observe from eq. (\ref{eq8}) that the transformations for the
coefficient $b^+$, $b^-$ are consistent with the relativistic
composition of the  velocities.

\section{Moment analysis and diffusivity transformation}
\label{sec4}

The statistical properties  of the stochastic process considered
in the moving reference system $\Sigma$ can be conveniently approached by considering the associated moment hierarchy
\begin{equation}
m_n(t)= \int_{-\infty}^\infty x^n \, p(x,t) \, d x =
m_n^+(t)+m_n^-(t) 
\label{eq11}
\end{equation}
where the partial moments 
\begin{equation}
m_n^\pm(t) = \int_{-\infty}^\infty x^n \, p^\pm(x,t) \, d x 
\label{eq12}
\end{equation}
satisfy the system of equations
\begin{eqnarray}
\frac{d m^+_n}{dt} & = & n \, b^+ \, m_{n-1}^+ - a^+ \, m_n^+ + a^- m_n^-
\nonumber \\
\frac{d m^-_n}{dt} & = & - n \, b^- \, m_{n-1}^- + a^+ \, m_n^+ - a^- m_n^-
\label{eq13}
\end{eqnarray}

For $n=0$ (zero-th order moment) eqs. (\ref{eq13}) reduce to:
\begin{equation}
\frac{d m_0^\pm}{dt} = \mp a^+ \, m^+_0 \pm a^-  \, m_0^- 
\end{equation}
In the long-term limit, (here the concept of long-term or asymptotic
property refers to  times larger than the characteristic
timescale characterizing the recombination dynamics between the
two probability waves $p^+$ and $p^-$ describing statistically the
Poisson-Kac process, and corresponds to timescales 
$t \gg \max \{ 1/a^+, 1/a^- \}$),
the zero-th order moments converge to steady values satisfying the
relation 
\begin{equation}
m_0^+  =\frac{a^-}{a^+} \, m_0^-
\label{eq14}
\end{equation}
Enforcing the consistency condition $m_0^++m_0^-=1$, one
thus obtains
\begin{equation}
m_0^\pm = \frac{a^\mp}{a^+ \, + \, a^-}  = \frac{1 \mp b^\prime_0 \, w/c^2}{2} 
\label{eq15}
\end{equation}
Next, consider the first-order moments, i.e., $n=1$, for which
eqs. (\ref{eq13}) provide
\begin{eqnarray}
\frac{d m_1^+}{d t} & = & b^+ \, m_0^+ - a^+ \, m_1^+ + a^- \, m_1^-
\nonumber \\
\frac{d m_1^-}{d t} & = & - b^- \, m_0^- + a^+ \, m_1^+ - a^- \, m_1^- 
\label{eq16}
\end{eqnarray}
The long-term  behavior of $m^\pm_1(t)$ is at most linear in time, i.e.,
\begin{equation}
m^\pm_1(t)= \mu_1^\pm \, t + \delta_1^\pm 
\label{eq17}
\end{equation}
Substituting these expressions into eqs. (\ref{eq16}), and equating
the coefficients of equal powers of $t^p$, $p=0,1$, one
obtains
the algebraic relations
\begin{equation}
a^+ \, \mu_1^+ - a^- \, \mu_1^- =0 
\label{eq18}
\end{equation}
and
\begin{eqnarray}
\mu_1^+ & = & b^+ \, m_0^+ - a^+ \, \delta_1^+ + a^- \, \delta_1^-
\nonumber \\
\mu_1^- & = & -b^- \, m_0^- + a^+ \, \delta_1^+ - a^- \, \delta_1^- 
\label{eq19}
\end{eqnarray}
Summing the two expressions (\ref{eq19}), a further
linear equation in $\mu_1^\pm$ is obtained, 
\begin{equation}
\mu_1^+ + \mu_1^- = b^+ \, m_0^+ - b^- \, m_0^- 
\label{eq20}
\end{equation}
Enforcing eq. (\ref{eq15}), it follows readily that
\begin{equation}
\mu_1= \mu_1^++ \mu_1^- = -w
\label{eq21}
\end{equation}
where $\mu_1$ is the effective velocity measured in the moving
reference system, $m_1(t) \sim \mu_1 \, t$. Eq. (\ref{eq21})
is physically straightforward, implying that the effective mean
velocity of the stochastic process measured in the moving system
is just the reverse velocity -$w$.
Eqs. (\ref{eq18}) and (\ref{eq20}) provide a linear system
in the two unknown $\mu_1^\pm$, yielding as a solution
\begin{equation}
\mu_1^\pm = \frac{ a^\mp \, (b^+ \, m_0^+ - b^- \, m_0^-)}{a^+ + a^-}  
\label{eq22}
\end{equation}

Finally, in order to derive dispersion properties, consider
the second-order moments $m_2^\pm$, which satisfy the
system of equations
\begin{eqnarray}
\frac{d m^+_2}{d t} & = & 2 \, b^+ \, m_1^+ - a^+ \, m_2^+ + a^- \, m_2^-
\nonumber \\
\frac{d m^-_2}{d t} & = & - 2 \, b^- \, m_1^- + a^+ \, m_2^+ - a^- \, m_2^-
\label{eq23}
\end{eqnarray}
In the long-term limit, $m_2^\pm(t)$ attains a quadratic expression
in $t$,
\begin{equation}
m_2^\pm(t) = \sigma_2^\pm \, t^2 + \zeta_2^\pm \, t + \eta_2^\pm \; .
\label{eq24}
\end{equation}
The substitution of eq. (\ref{eq24}) into eq. (\ref{eq23}) provides
the system of linear relations in the long-term coefficients
\begin{equation}
a^+ \, \sigma_2^+ - a^- \, \sigma_2^- =0 
\label{eq25}
\end{equation}
\begin{eqnarray}
2 \sigma_2^+  & = & 2 b^+ \, \mu_1^+ - a^+ \, \zeta_2^+ + a^- \, \zeta_2^-
\nonumber \\
2 \sigma_2^- & = & - 2 b^- \, \mu_1^- + a^+ \, \zeta_2^+ - a^- \, \zeta_2^- 
\label{eq26}
\end{eqnarray}
and,
\begin{eqnarray}
\zeta_2^+ & = & 2 b^+ \, \delta_1^+ - a^+ \, \eta_2^+ + a^- \, \eta_2^-
\nonumber \\
\zeta_2^- & = & -2 b^- \, \delta_1^- + a^+ \, \eta_2^+ - a^- \, \eta_2^- 
\label{eq27}
\end{eqnarray}
Summing together the two expressions (\ref{eq26}), and using eq.
(\ref{eq25}), a system of two linear equations for $\sigma_2^\pm$
is obtained, the solution of which is
\begin{equation}
\sigma_2^\pm = \frac{a^\mp}{a^+ + a^-} \,
 \left ( b^+ \mu_1^+ - b^- \, \mu_1^- \right ) 
\label{eq28}
\end{equation}

We are mainly interested in the scaling  of the mean square
displacement $\sigma^2(t)$,
\begin{eqnarray}
\sigma^2(t)  & = & m_2(t) - [m_1(t)]^2 \nonumber \\
& = & m_2^+(t) + m_2^-(t) - \left [ m_1^+(t) + m_1^-(t) \right ]^2 
\label{eq29}
\end{eqnarray}
Asymptotically, i.e., for time-scales in which the recombination
process between backward and forward probability waves has
reached a stationary behavior, the expression for $\sigma^2(t)$
attains the form
\begin{equation}
\sigma^2(t) = A_2 \, t^2 + \Sigma(v) \, t + \Delta 
\label{eq30}
\end{equation}
where
\begin{eqnarray}
A_2 & = & \sigma_2^+ + \sigma_2  -(\mu_1^+ + \mu_1^-)^2 \nonumber \\
\Sigma(v) & = & (\zeta_2^+ + \zeta_2^-) - 2 (\mu_1^++\mu_1^-) (\delta_1^+
+ \delta_1^-) \label{eq31} \\
\Delta & = & (\eta_2^+ + \eta_2^-) - (\delta_1^+ + \delta_1^-)^2 
\nonumber 
\end{eqnarray}
From the expressions obtained for $\sigma^\pm_2$ and $\mu_1^\pm$ it follows
identically that $A_2=0$, so that, as expected,
\begin{equation}
\sigma^2(t) \sim \Sigma(w) \, t 
\label{eq32}
\end{equation}
Summing together the two equations (\ref{eq27}) one obtains
\begin{equation}
\zeta^+_2+\zeta_2^- = 2  \left (  b^+ \, \delta_1^+ - b^- \, \delta_1^- \right ) \; .
\label{eq33}
\end{equation}
This equation, together with eq. (\ref{eq20}) provide, after some algebra,
 the following expression for $\Sigma(w)$
\begin{equation}
\Sigma(w) = 2 \, \left ( b^+ - \mu_1^+-\mu_1^- \right )  \, 
\left ( \frac{\mu_1^- + b^- \, m_0^-}{a^+}  \right ) 
\label{eq34}
\end{equation}
Substituting the expressions for $\mu_1^\pm$, $m_0^-$, $b^\pm$ and $a^+$
into eq. (\ref{eq34}), one finally arrives to the
compact expression
\begin{equation}
\frac{\Sigma(w)}{2} = \frac{(b^\prime_0)^2}{2 \, a_0^\prime} \, \left ( 1 - \frac{w^2}{c^2}
\right )^{3/2} 
\label{eq35}
\end{equation}
But $\Sigma(v)/2$ is  the diffusion coefficient $D$ in the moving
reference frame, while, from eq. (\ref{eq3}), $(b^\prime_0)^2/2 a_0^\prime$ 
is the rest diffusion coefficient $D_0$.
Consequently, eq. (\ref{eq35}) can be expressed in a compact way as
\begin{equation}
D = D_0  \left ( 1 - \frac{w^2}{c^2}
\right )^{3/2} = D_0 \gamma^{-3}(w) 
\label{eq36}
\end{equation}
which is the transformation connecting the diffusion coefficients
in the two inertial frames.

\section{Extension to higher  spatial dimensions}
\label{sec5}

The analysis developed in the previous Section is extended to higher
spatial dimensions. Without loss of generality,
  two spatial coordinates are considered.

\subsection{Kolesnik-Kac stochastic process}
\label{sec5_1}

As a two-dimensional model of a Poisson-Kac stochastic process we
consider that proposed by Kolesnik and Turbin \cite{kolesnik} and Kolesnik \cite{kolesnik2}, and therefore referred to
as the {\em Kolesnik-Kac model}. It is a particular case
of the class of Generalized Poisson-Kac processes introduced in
\cite{giona_GPK}. 

Consider a system of $N>2$ uniform velocity vectors
\begin{equation}
{\bf b}^{\prime,(\alpha)}= b_0^\prime  \, \left (
\cos(2\pi \alpha/N),\sin(2 \pi \alpha/N) \right )
= (b_x^ {\prime,(\alpha)},b_y^{\prime,(\alpha)})
\; , \qquad \alpha=0,\dots,N-1
\label{eq5_1}
\end{equation}
where $b_0$ is the reference velocity, since $|{\bf b}^{\prime,(\alpha)}|=b_0$
for any $\alpha$, and
the stochastic process
\begin{equation}
d {\bf x}^\prime(t^\prime) = {\bf w}(t^\prime) \, d t^\prime 
\label{eq5_2}
\end{equation}
${\bf x}^\prime =(x^\prime,y^\prime)$, where ${\bf w}(t^\prime)$
is a stochastic velocity vector that change its value,
amongs the $N$ possible states ${\bf b}^{\prime'(\alpha)}$ in a equiprobable
way, at random time intervals characterized by an exponential
distribution defined by the transition rate $a_0^\prime>0$.
The mathematical properties of this process have been
studied by Kolesnik \cite{kolesnik}.

In the stationary system $\Sigma^\prime$ (space-time coordinates
$x^\prime$, $y^\prime$, $t^\prime$), a system of $N$ partial
probability density functions 
$p^{\prime, (\alpha)}(x^\prime,y^\prime,t^\prime)$, $\alpha=0,\dots,N$, fully
describes  the statistical properties  of the Kolesnik-Kac model, and the
overall 
probability density function of the process is $p^\prime(x^\prime,y^\prime,t^\prime)= \sum_{\alpha=0}^{N-1} p^{\prime, (\alpha)}(x^\prime,y^\prime,t^\prime)$.

In the stationary frame the partial probabilities satisfy the system
of hyperbolic equations
\begin{equation}
\frac{\partial p^{\prime,(\alpha)}}{\partial t^\prime}= - b_x^{\prime,(\alpha)}
\, \frac{\partial p^{\prime,(\alpha)}}{\partial x^\prime}
- b_y^{\prime,(\alpha)}
\, \frac{\partial p^{\prime,(\alpha)}}{\partial y^\prime}
- a_0^\prime \, p^{\prime,(\alpha)} + \frac{a_0^\prime}{N} \sum_{\beta=0}^{N-1} 
p^{\prime,(\beta)} 
\label{eq5_3}
\end{equation}
$\alpha=0,\dots,N-1$.
In an inertial frame $\Sigma$ (space-time coordinates $x$, $y$, $t$) moving 
with respect to $\Sigma^\prime$ at constant velocity $w \, {\bf e}_x$
 along the $x$-axis,
the process is characterized by the $N$ partial probability densities
$p^{(\alpha)}(x,y,t)$, $\alpha=0,\dots,N-1$, 
defined as
\begin{equation}
p^{(\alpha)}=  \gamma(w) \, \kappa(\alpha) \, p^{\prime, (\alpha)} \; ,
\qquad \alpha=0,\dots,N-1 \, ,
\label{eq5_4}
\end{equation}
where 
\begin{equation}
\kappa(\alpha)= 1- \frac{b_x^{\prime,(\alpha)} \, w}{c^2}
\label{eq5_5}
\end{equation}
The balance equation for the partial probability system $\{p^{(\alpha)} \}_{\alpha=0}^{N-1}$ is obtained from eq. (\ref{eq5_3}) enforcing  
the Lorentz transform (\ref{eq4}) and $y=y^\prime$, leading to
\begin{equation}
\frac{\partial p^{(\alpha)}}{\partial t}= - b_x^{(\alpha)} \,
\frac{\partial p^{(\alpha)}}{\partial x} - b_y^{(\alpha)} \,
\frac{\partial p^{(\alpha)}}{\partial y} - a^{(\alpha)} \, p^{(\alpha)}
+ \frac{1}{N} \sum_{\beta=0}^{N-1} a^{(\beta)} \, p^{(\beta)} 
\label{eq5_6}
\end{equation}
$\alpha=0,\dots,N-1$. The coefficients entering  these equations
are defined by
\begin{equation}
b_x^{(\alpha)} = \frac{b_x^{\prime,(\alpha)}-w}{\kappa(\alpha)}
\; , \qquad 
b_y^{(\alpha)} = \frac{b_y^{\prime,(\alpha)}}{ \gamma(w) \, \kappa(\alpha)}
\; , \qquad
a^{(\alpha)} = \frac{a_0^\prime}{\gamma(w) \, \kappa(\alpha)}
\label{eq5_7}
\end{equation}
$\alpha=0,\dots,N-1$.

\subsection{Moment hierarchy}
\label{sec5_2}

As in the   one-dimensional  spatial case  previously analyzed,
 moment analysis  provides the
simplest tool to extract the statistical properties associated 
with eqs. (\ref{eq5_6}). Let
\begin{equation}
m_{m,n}^{(\alpha)}(t) = \int_{-\infty}^\infty \int_{-\infty}^\infty
x^m \, y^n \, p^{(\alpha)}(x,y,t) \, dx \, dy 
\label{eq5_8}
\end{equation}
$m,n=0,1,\dots$, $\alpha=0,\dots,N-1$ be  the $(m,n)$-th partial moment.
The global moment hierarchy $\{m_{m,n}(t)\}_{m,n=0}^\infty$ defined
with respect to $p(x,y,t)$ is
readily given by the sum over $\alpha$ of the $(m,n)$-th partial
moments.
From the definition (\ref{eq5_8}) and from eqs. (\ref{eq5_6}),
it follows
 that the partial  moments satisfy the linear system of differential
equations
\begin{equation}
\frac{d m_{m,n}^{(\alpha)}}{d t}= m \, b_x^{(\alpha)} m_{m-1,n}^{(\alpha)}
+ n \, b_y^{(\alpha)} \, m_{m,n-1}^{(\alpha)} -a^{(\alpha)} \,
m_{m,n}^{(\alpha)} + \frac{1}{N} \sum_{\beta=0}^{N-1} a^{(\beta)}
\, m_{m,n}^{(\beta)}  
\label{eq5_9}
\end{equation}
where the
 regularity condition at infinity for $p^{(\alpha)}(x,y,t)$ has been
enforced.

\subsection{Zero-th and first-order moments}
\label{sec5_3}

To begin with, consider the zero-th order moments $m_{0,0}^{(\alpha)}$,
satisfying the system of equations
\begin{equation}
\frac{d m_{0,0}^{(\alpha)}}{d t} = - a^{(\alpha)} \, m_{0,0}^{(\alpha)}
+ \frac{1}{N} \sum_{\beta=0}^{N-1} a^{(\beta)}
\, m_{0,0}^{(\beta)}  
\label{eq5_10}
\end{equation}
$\alpha=0,\dots,N-1$. In the long-term (asymptotic) limit, i.e.,
once the recombination amongst the partial waves has reached a steady-state,
$d m_{0,0}^{(\alpha)}/d t=0$, which implies $m_{0,0}^{(\alpha)}= C^\prime/a^{(\alpha)}= C \kappa(\alpha)$, where the constant $C$ is determined by
the probabilistic consistency condition $\sum_{\alpha=0}^{N-1} m_{0,0}^{(\alpha)}=1$, leading to the expression
\begin{equation}
m_{0,0}^{(\alpha)} = \frac{\kappa(\alpha)}{N}
\; , \qquad  \alpha=0,\dots,N-1 
\label{eq5_11}
\end{equation}

Next, consider the  first-order  moments. For $m=1$, $n=0$, eqs. (\ref{eq5_9})
become 
\begin{equation}
\frac{d m_{1,0}^{(\alpha)}}{d t} = b_x^{(\alpha)} \, m_{0,0}^{(\alpha)}
- a^{(\alpha)} \, m_{1,0}^{(\alpha)}
+ \frac{1}{N} \sum_{\beta=0}^{N-1} a^{(\beta)}
\, m_{1,0}^{(\beta)}  
\label{eq5_12}
\end{equation}
$\alpha=0,\dots,N-1$. In the long-term limit eqs. (\ref{eq5_11})
become a non-homogeneous linear system driven by
a constant contribution $b_x^{(\alpha)} \, m_{0,0}^{(\alpha)}$,
where $m_{0,0}^{(\alpha)}$ is given by eq. (\ref{eq5_11}).
Consequently, the asymptotic solution for $m_{1,0}^{(\alpha)}$
grows linearly in time, i.e.,
\begin{equation}
m_{1,0}^{(\alpha)}= \mu_x^{(\alpha)} \, t + \delta_x^{(\alpha)}
\label{eq5_13}
\end{equation}
$\alpha=0,\dots,N-1$. Substituting, these expressions into eqs. (\ref{eq5_12})
and equating the coefficients of equal powers of $t$, one obtains the
system of relations for the asymptotic coefficients $ \mu_x^{(\alpha)}$,
$\delta_x^{(\alpha)}$
\begin{equation}
a^{(\alpha)} \, \mu_x^{(\alpha)} = \frac{1}{N} \sum_{\beta=0}^{N-1} a^{(\beta)}
\mu_x^{(\beta)}  
\label{eq5_14}
\end{equation}
\begin{equation}
\mu_x^{(\alpha)} = b_x^{(\alpha)} \, m_{0,0}^{(\alpha)} 
- a^{(\alpha)} \, \delta_x^{(\alpha)}
+ \frac{1}{N} \sum_{\beta=0}^{N-1} a^{(\beta)}
\, \delta_x^{(\beta)}  
\label{eq5_15}
\end{equation}
$\alpha=0,\dots,N-1$, where $m_{0,0}^{(\alpha)}$ equals its long-term
expression (\ref{eq5_11}).

Summing over $\alpha$ in eq. (\ref{eq5_15})
the expression for the effective velocity $V_x$,
\begin{equation}
m_{1,0}(t) \sim V_x \, t 
\label{eq5_16}
\end{equation}
is obtained
\begin{equation}
V_x = \sum_{\alpha=0}^{N-1} \mu_x^{(\alpha)} = \sum_{\alpha=0}^{N-1}
b_x^{(\alpha)} \, m_{0,0}^{(\alpha)} = - w 
\label{eq5_17}
\end{equation} 
and from  eqs. (\ref{eq5_14}), (\ref{eq5_17}) it follows that
the expression for $\mu_x^{(\alpha)}$ takes the form
\begin{equation}
\mu_x^{(\alpha)} = - \frac{w \, \kappa(\alpha)}{N} 
\label{eq5_18}
\end{equation}
$\alpha=0,\dots,N-1$.  From eqs. (\ref{eq5_15}) one also
obtains the functional form of $\delta_x^{(\alpha)}$,
that plays a central role in the estimate of diffusional properties.
Let $C_x=\sum_{\alpha=0}^{N_1}
a^{(\alpha)} \, \delta_x^{(\alpha)}$, eqs. (\ref{eq5_15})
can be rewitten as
\begin{equation}
\delta_x^{(\alpha)} = \frac{1}{a^{(\alpha)}} \, \left [
 b_x^{(\alpha)} \, m_{0,0}^{(\alpha)} - \mu_x^{(\alpha)} + \frac{C_x}{N}
\right ]
\label{eq5_19}
\end{equation}
$\alpha=0,\dots,N-1$. Substituting the long-term expressions 
(\ref{eq5_11}), (\ref{eq5_18}) one finally arrive to
\begin{equation}
\delta_x^{(\alpha)} = \frac{\gamma(w)}{N \, a_0} \, \left ( b_x^{\prime, (\alpha)} - w \right ) \,
\kappa(\alpha) + \frac{ w \, \gamma(w)}{N \, a_0} \, \kappa^2(\alpha) +
\frac{ \gamma(w) \, C_x}{ N \, a_0} \, \kappa(\alpha) 
\label{eq5_20}
\end{equation}
$\alpha=0,\dots,N-1$.  The expressions for $\delta_x^{(\alpha)}$ contain
the unknown constant $C_x$. As we shall see in the next paragraph,
 this constant
is totally immaterial in the estimate of the diffusional properties.

Finally, consider the other family of first-order partial moments $m_{0,1}^{(\alpha)}$. The analysis is completely specular to that developed above for
$m_{1,0}^{(\alpha)}$ so that solely the final results are reported.

In the long-term limit, $m_{0,1}^{(\alpha)}(t)$ attains a linear expression
analogous to  eq. (\ref{eq5_13}), namely,
\begin{equation}
m_{0,1}^{(\alpha)}(t) = \mu_y^{(\alpha)} \, t + \delta_y^{(\alpha)}
\label{eq5_21}
\end{equation}
$\alpha=0,\dots,N-1$. From the balance equations it follows that
\begin{equation}
\mu_y^{(\alpha)}=0 
\label{eq5_22}
\end{equation}
$\alpha=0,\dots,N-1$, which implies for $V_y$, $m_{0,1}(t) \sim V_y t$ that
\begin{equation}
V_y=0
\label{eq5_23}
\end{equation}
As it regards the factors $\delta_y^{(\alpha)}$ entering the
asymptotic functional form of these first-order partial moments,
one obtains
\begin{equation}
\delta_y^{(\alpha)}  =  \frac{1}{a^{(\alpha)}} \, \left [ 
b_y^{(\alpha)} \, m_{0,0}^{(\alpha)} + \frac{C_y}{N} \right ]  
 =  \frac{1}{N \, a_0} \beta_y^{(\alpha)} \, \kappa(\alpha) + 
\frac{\gamma(w) \, C_y}{N \, a_0} \, \kappa(\alpha)  
\label{eq5_24}
\end{equation}
$\alpha=0,\dots,N-1$. In deriving eqs. (\ref{eq5_24}), eqs. (\ref{eq5_22})
have been used. The coefficients $\delta_y^{(\alpha)}$ contain
the constant $C_y= \sum_{\alpha=0}^{N-1} a^{(\alpha)} \, \delta_y^{(\alpha)}$,
which is immaterial in the estimate of the long-term 
diffusional properties, that
are addressed in the next paragraph.

\subsection{Second-order moments and diffusion tensor}
\label{sec5_4}

To begin with, consider $m_{2,0}^{(\alpha)}$, i.e., $m=2$ and $n=0$.
Eqs. (\ref{eq5_9}) specialize into
\begin{equation}
\frac{d m_{2,0}^{(\alpha)}}{d t} = 2 \, b_x^{(\alpha)} \, m_{1,0}^{(\alpha)}
- a^{(\alpha)} \, m_{2,0}^{(\alpha)}
+ \frac{1}{N} \sum_{\beta=0}^{N-1} a^{(\beta)}
\, m_{2,0}^{(\beta)}  
\label{eq5_25}
\end{equation}
$\alpha=0,\dots,N-1$. In the long-term limit, since the
forcing term $ b_x^{(\alpha)} \, m_{1,0}^{(\alpha)}$ is linear in time, 
$m_{2,0}^{(\alpha)}(t)$ should be, at most, a quadratic function of $t$
\begin{equation}
m_{2,0}^{(\alpha)}(t) = K_{x,x}^{(\alpha)} \, t^2 + g_{x,x}^{(\alpha)}
\, t + \zeta_{x,x}^{(\alpha)} 
\label{eq5_26}
\end{equation}
$\alpha=0,\dots,N-1$. Substituted into eq. (\ref{eq5_25}), it
provides the following relations amongst the expansion coefficients
\begin{equation}
a^{(\alpha)} \, K_{x,x}^{(\alpha)} = \frac{1}{N} \sum_{\beta=0}^{N-1}
a^{(\beta)} \, K_{x,x}^{(\beta)}  
\label{eq5_27}
\end{equation}
\begin{equation}
2 K_{x,x}^{(\alpha)} = 2 b_x^{(\alpha)} \, \mu_x^{(\alpha)} -
a^{(\alpha)} \, g_{x,x}^{(\alpha)} + \frac{1}{N} \sum_{\beta=0}^{N-1}
a^{(\beta)} \, g_{x,x}^{(\beta)} 
\label{eq5_28}
\end{equation}
\begin{equation}
g_{x,x}^{(\alpha)} = 2 b_x^{(\alpha)} \, \delta_x^{(\alpha)}
- a^{(\alpha)} \, \zeta_{x,x}^{(\alpha)} +
\frac{1}{N} \sum_{\beta=0}^{N-1}
a^{(\beta)} \, \zeta_{x,x}^{(\beta)} 
\label{eq5_29}
\end{equation}
$\alpha=0,\dots,N-1$.
Summing  over $\alpha$ in eq. (\ref{eq5_28}), and enforcing
eqs.  (\ref{eq5_18}), it follows that
\begin{equation}
\sum_{\alpha=0}^{N-1} K_{x,x}^{(\alpha)} = \sum_{\alpha=0}^{N-1}
b_x^{\alpha} \, \mu_x^{(\alpha)} = w^2 
\label{eq5_30}
\end{equation}
that,  together with eqs. (\ref{eq5_27}),
 provides for $K_{x,x}^{(\alpha)}$
the expression
\begin{equation}
K_{x,x}^{(\alpha)} = \frac{w^2 \, \kappa(\alpha)}{N} 
\label{eq5_31}
\end{equation}
$\alpha=0,\dots,N-1$.
The variance $\sigma_{x,x}^2(t)$ along the $x$-coordinate is
given by
\begin{equation}
\sigma_{x,x}^2(t) = \sum_{\alpha=0}^{N-1} m_{2,0}^{(\alpha)}(t) - \left [ \sum_{\alpha=0}^{N-1} m_{1,0}^{(\alpha)}(t) \right ]^2 
\label{eq5_32}
\end{equation}
Substituting the long-term expressions and enforcing eqs. (\ref{eq5_18}) and
(\ref{eq5_31}), it follows that $\sigma_{x,x}^2(t)$ is asymptotically
a linear function of time
\begin{equation}
\sigma_{x,x}^2(t) \sim 2 \,  D_{x,x} \, t 
\label{eq5_33}
\end{equation}
where the diffusivity along $x$ is given by the expression
\begin{equation}
2 \, D_{x,x}  = \sum_{\alpha=0}^{N-1} g_{x,x}^{(\alpha)}
+ 2 \, w \, \sum_{\alpha=0}^{N-1} \delta_x^{(\alpha)} \; .
\label{eq5_34}
\end{equation}
The first sum entering eq. (\ref{eq5_34}) can be explicited by
summing eqs.  (\ref{eq5_29}) over $\alpha$, providing the
compact expression for $D_{x,x}$
\begin{equation}
D_{x,x}= \sum_{\alpha=0}^{N-1}  \left ( b_x^{(\alpha)} + w \right )
\, \delta_x^{(\alpha)} 
\label{eq5_35}
\end{equation}
where
\begin{equation}
b_x^{(\alpha)} + w= \frac{ b_x^{\prime,(\alpha)}}{\gamma^2(w) \, \kappa(\alpha)}
\label{eq5_36}
\end{equation}
$\alpha=0,\dots,N-1$. From the latter expression, and from the
expression (\ref{eq5_35}), it follows that the term containg the unknown
constant $C_x$ in the expression (\ref{eq5_19}) gives a vanishing
contribution in the estimate of $D_{x,x}$.
Therefore,
\begin{eqnarray}
D_{x,x} & = &  \frac{1}{\gamma^2(w)} \sum_{\alpha=0}^{N-1}
\frac{b_x^{\prime,(\alpha)} }{\kappa(\alpha)}
\left [ \frac{\gamma(w)}{N \, a_0} \left ( b_x^{\prime,(\alpha)} - w \right )
\, \kappa(\alpha) + \frac{ w \, \gamma(w)}{N \, a_0} \kappa^2(\alpha) \right ]
\nonumber \\
& = & \frac{1}{\gamma(w) \, N \, a_0^\prime} \left [
\sum_{\alpha=0}^{N-1} b_x^{\prime,(\alpha)} \,  \left ( b_x^{\prime,(\alpha)} - w \right ) + w \, \sum_{\alpha=0}^{N-1}  b_x^{\prime,(\alpha)} \, \kappa(\alpha) \right ] \nonumber \\
& = &  \frac{1}{\gamma(w) \, N \, a_0^\prime}  \left ( 1- \frac{w^2}{c^2} \right )
\, \sum_{\alpha=0}^{N-1} \left [ b_x^{\prime,(\alpha)} \right ]^2
= \frac{(b_0)^2}{2 \, a_0} \, \gamma^{-3}(w) 
\label{eq5_37}
\end{eqnarray}
where we have used the identities $\sum_{\alpha=0}^{N-1} b^{\prime,(\alpha)}=0$,
$\sum_{\alpha=0}^{N-1} \left [ b_x^{\prime,(\alpha)} \right ]^2 = (b_0)^2 N/2$.
Equation (\ref{eq5_37}) implies the transformation
of  the  diffusion coefficient $D_{x,x}$
parallel to the frame-velocity direction
\begin{equation}
D_{x,x}= D_\parallel = D_0 \, \gamma^{-3}(w) 
\label{eq5_38}
\end{equation}
consistently with the corresponding expression  (\ref{eq36}) derived for
the   one-dimensional   spatial Poisson-Kac process.

Consider now the other  family $m_{0,2}^{(\alpha)}$, corresponding to
$m=0$ and $n=2$. Details are skipped as the algebra is identical
to the $m_{2,0}^{(\alpha)}$-case. In the
long-term regime, $m_{0,2}^{(\alpha)}$ is quadratic in time
\begin{equation}
m_{0,2}^{(\alpha)} = K_{y,y}^{(\alpha)} \, t^2 + g_{y,y}^{(\alpha)} \, t
+ \zeta_{y,y}^{(\alpha)} 
\label{eq5_39}
\end{equation}
$\alpha=0,\dots,N-1$. 
From the moment balance  equation one obtains
\begin{equation}
K_{y,y}^{(\alpha)} =0
\label{eq5_40}
\end{equation}
$\alpha=0,\dots,N-1$. As  regards $\sigma_{y,y}^2(t)= m_{0,2}(t) -\left [m_{0,1}(t) \right ]^2$ one obtains
\begin{equation}
\sigma_{y,y}^2(t) \sim 2 \, D_{y,y} \, t 
\label{eq5_41}
\end{equation}
where the diffusion coefficient $D_{y,y}$ is given by
\begin{equation}
 2 \, D_{y,y}= \sum_{\alpha=0}^{N-1} g_{y,y}^{(\alpha)}
= 2 \, \sum_{\alpha=0}^{N-1} b_y^{(\alpha)} \, \delta_y^{(\alpha)} 
\label{eq5_42}
\end{equation}
Substituting the expressions (\ref{eq5_7}) and (\ref{eq5_24})
for $b_y^{(\alpha)}$ and $\delta_y^{(\alpha)}$ into
eq. (\ref{eq5_42}) one gets
\begin{eqnarray}
D_{y,y} & = & \sum_{\alpha=0}^{N-1} \frac{b_y^{\prime, (\alpha)}}{\gamma(w) \,
\kappa(\alpha)} \, \left [ \frac{1}{N \, a_0}  \, b_y^{\prime, (\alpha)}
\, \kappa(\alpha) + \frac{\gamma(w) \, C_y}{N \, a_0} \, \kappa(\alpha)
\right ] \nonumber \\
& = & \frac{1}{\gamma(w) \, N \, a_0^\prime} 
 \sum_{\alpha=0}^{N-1} \left [ b_y^{\prime,(\alpha)} \right ]^2
= \frac{(b_0)^2}{ 2 \, a_0} \, \gamma^{-1}(w)
\label{eq5_43}
\end{eqnarray}
which implies that
\begin{equation}
D_{y,y}= D_\perp = D_0 \gamma^{-1}(w)= D_0 \, \sqrt{ 1- \frac{w^2}{c^2}} 
\label{eq5_44}
\end{equation}
i.e., the effective
 diffusion coefficient perpendicular to the frame direction of motion
is proportional to 
the reciprocal of the Lorentz factor.

Finally, consider the mixed moments $m_{1,1}^{(\alpha)}$, i.e.,
$m=n=1$. Also in this case, in the long-term
\begin{equation}
m_{1,1}^{(\alpha)} = K_{x,y}^{(\alpha)} \, t^2 + g_{x,y}^{(\alpha)} \, t
+ \zeta_{x,y}^{(\alpha)} 
\label{eq5_45}
\end{equation}
$\alpha=0,\dots,N-1$.
From the moment balance equation it follows that
\begin{equation}
K_{x,y}^{(\alpha)} = 0 
\label{eq5_46}
\end{equation}
$\alpha=0,\dots,N-1$, and
\begin{equation}
\sigma_{x,y}^2(t) = m_{1,1}(t)- m_{1,0}(t) \, m_{0,1}(t)
\sim 2 D_{x,y} \, t 
\label{eq5_47}
\end{equation}
where
\begin{eqnarray}
2 \, D_{x,y}  & = & \sum_{\alpha=0}^{N-1} g_{x,y}^{(\alpha)} -
\sum_{\alpha=0}^{N-1} \mu_x^{(\alpha)} \, \sum_{\beta=0}^{N-1} \delta_x^{(\beta)}  \nonumber \\
& = & \sum_{\alpha=0}^{N-1} \left ( b_x^{(\alpha)}+ w \right )
\, \delta_y^{(\alpha)} + \sum_{\alpha=0}^{N-1} b_y^{(\alpha)} \, \delta_x^{(\alpha)}  
\label{eq5_48}
\end{eqnarray}
Using the expressions previously derived for the quantities
entering eq. (\ref{eq5_48}), one derives after some algebra that
\begin{equation}
D_{x,y}=0 
\label{eq5_49}
\end{equation}
which completes the diffusional analysis of the Kolesnik-Kac process.

\section{Scaling analysis}
\label{sec6}

The transformations of the diffusion coefficient in inertial
systems can be physically interpreted  on the basis
of time-dilation and length-contraction phenomena
using the classical Einstein scaling for the diffusion coefficient.
The analysis is  not mathematically rigorous and it is  aimed
at highlighting the physical origin of the diffusivity transformations,
at least in these simple cases.
The diffusivity
$D^\prime$ is proportional to the ratio of the mean square
displacement 
$\langle( \Delta x^\prime)^2 \rangle$ divided by the time-scale
 $\Delta t^\prime$
\begin{equation}
D^\prime = \frac{1}{2} \, \frac{\langle ( \Delta x^\prime)^2 \rangle}{\Delta t^\prime} = D_0
\label{eq6_1}
\end{equation}
which is the Einstein equation for diffusion.
Similarly, in the reference frame  $\Sigma$, the diffusion coefficient is
given by
\begin{equation}
D = \frac{1}{2} \, \frac{\langle (\Delta x)^2 \rangle}{\Delta t}  
\label{eq6_2}
\end{equation}
But $\Delta t$, and $\Delta x$ are related to $\Delta x^\prime$ and $\Delta t$
via the Lorentz transformation, which  implies length contraction
\begin{equation}
\Delta x = \Delta x^\prime \, \sqrt{1-\frac{w^2}{c^2}} = \gamma^{-1}(w) \,
\Delta x^\prime 
\label{eq6_3}
\end{equation}
and time dilation
\begin{equation}
\Delta t =  \frac{\Delta t^\prime}{\sqrt{1-\frac{w^2}{c^2}}} = \gamma(w)
 \, \Delta
t^\prime 
\label{eq6_4}
\end{equation}
Therefore,
\begin{equation}
\langle (\Delta x)^2 \rangle =  \gamma^{-2}(w) \, 
\langle (\Delta x^\prime)^2 \rangle 
\label{eq6_3bis}
\end{equation}
Substituting eqs. (\ref{eq6_4}) and (\ref{eq6_3bis}) into
eq. (\ref{eq6_2}), eq. (\ref{eq36}) follows. Therefore, the scaling of
$D$ with the third power of $\gamma^{-1}(w)$ can be viewed
as a direct consequence of the space-time contraction/dilation
properties of the Lorenz transformation.

In higher-dimensional spaces,
\begin{equation}
D_{x_i,x_i} = 
\frac{1}{2} \, \frac{\langle (\Delta x_i)^2 \rangle}{\Delta t^\prime} 
\label{eq6_5}
\end{equation}
$i=1,\dots d$, where $d=2,3$, two cases sould be considered separately
depending whether $x_i$ is a spatial coordinate  in the
direction of the frame velocity or not.
In the first case, namely for a frame velocity directed along $x_i$,
$D_{x_i,x_i}=D_\parallel$, $\langle (\Delta x_i)^2 \rangle$ satisfies
eq. (\ref{eq6_3bis}) and eq. (\ref{eq5_38}) is recovered for
$D_\parallel$. Conversely, if  $x_i$ is a coordinate
in a direction orthogonal
to the frame velocity, $D_{x_i,x_i}=D_\perp$,
 then $ \langle ( \Delta x_i)^2 \rangle= \langle (\Delta x_i^\prime)^2 \rangle$, so that solely
time dilation contributes to $D_\perp$, returning
$D_\perp=D_0 \gamma^{-1}(w)$ as derived in  previous 
Section.
For generic stochastic processes, scaling analysis
is not sufficient to provide the expression for the complete  transformation
of the tensor diffusivity. This analysis is developed in Section
\ref{sec9}.

\section{Numerical examples}
\label{sec8}

It is useful to illustrate the main results developed in
the previous Sections with the aid of numerical
examples of stochatic dynamics.
Throughout this Section, a normalized light velocity is
assumed, i.e., $c=1$.

\begin{figure}
\includegraphics[width=12cm]{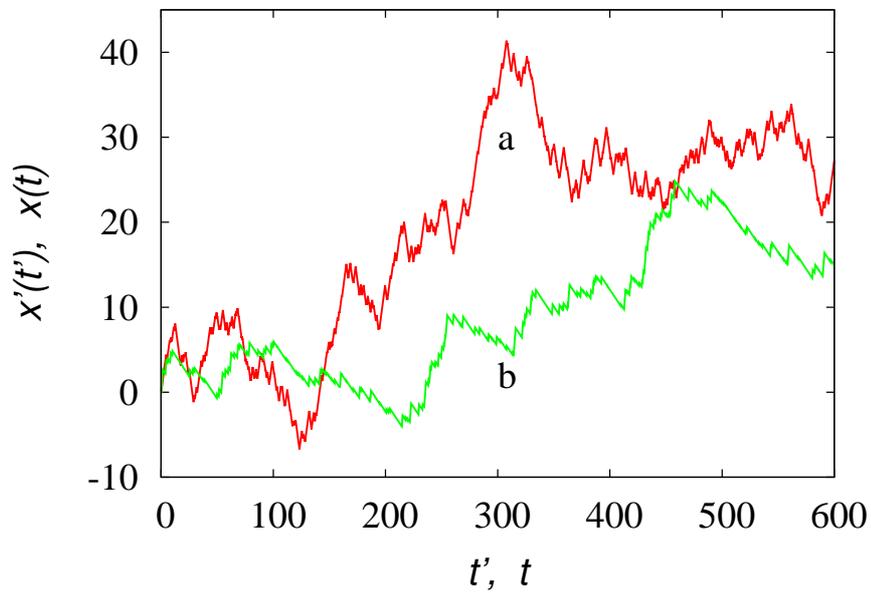}
\caption{Graph of the evolution of a realization
of a Poisson-Kac process ($b_0^\prime=c=1$, $a_0^\prime=1$) in
the rest frame $\Sigma^\prime$ (line a, i.e., $x^\prime(t^\prime$)),
and in the inertial system $\Sigma$ moving with constant velocity $w/c=0.8$
with respect to $\Sigma^\prime$
(line b, i.e., $x(t)$).}
\label{Fig_1}
\end{figure}

\begin{figure}[htb]
\begin{center}
\includegraphics[width=8.5cm]{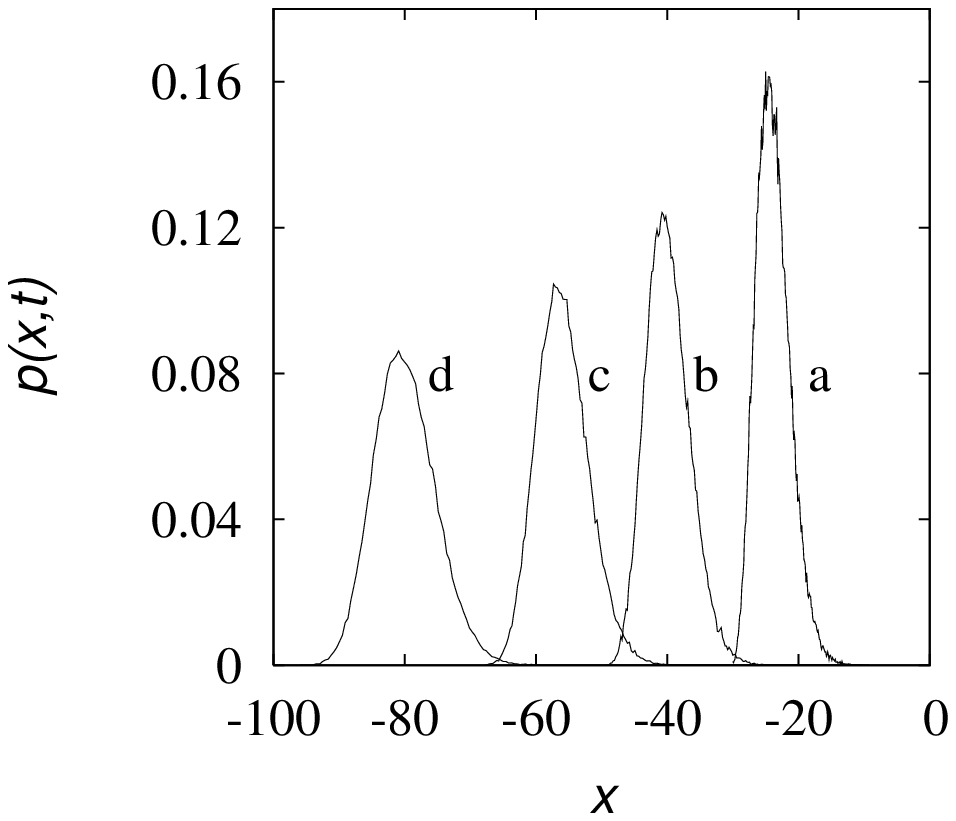}
\hspace{-2.2cm} {\Large (a)}
\includegraphics[width=8.5cm]{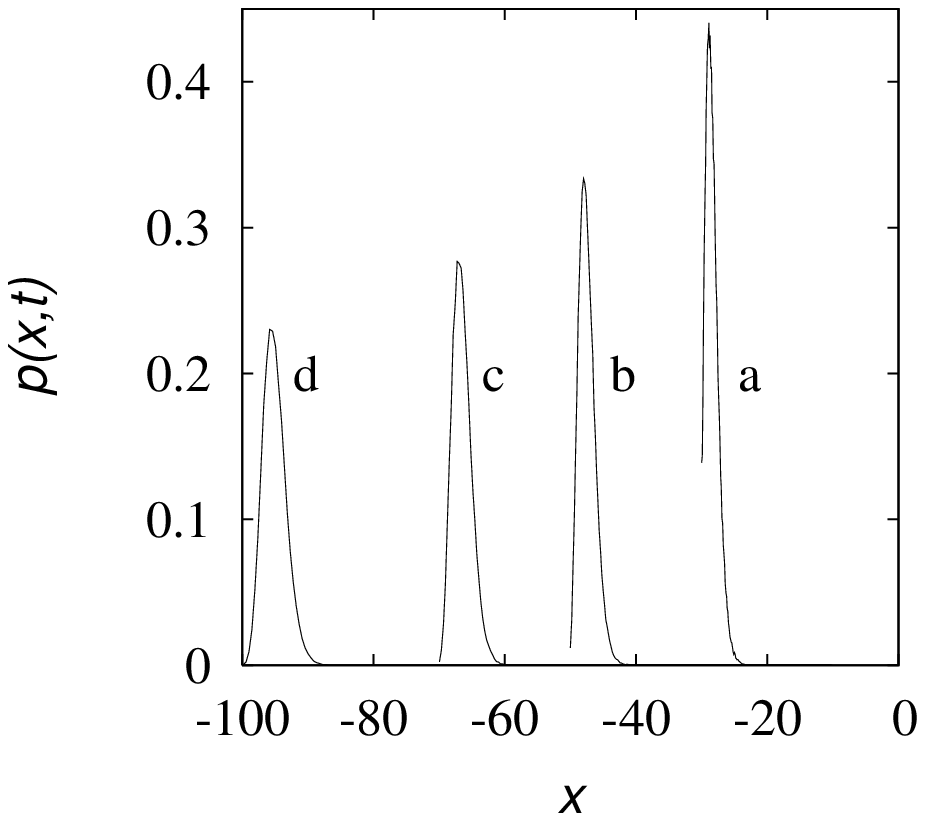}
\hspace{-2.2cm} {\Large (b)}
\end{center}
\caption{Overall probability density function $p(x,t)$ vs $x$
of a Poisson-Kac process ($b_0^\prime=c=1$, $a_0^\prime=1$) 
in the frame $\Sigma$
 at several  time instants. Panel (a)  refers to $w/c=0.8$,
panel (b)  to $w/c=0.95$. Lines (a) to (d) refer to $t=30,\, 50,\,70$ and 
$100$, respectively.}
\label{Fig_2}
\end{figure}

To begin with, consider the  one-dimensional  spatial (1d, for short)
 Poisson-Kac process eq. (\ref{eq1}).
Set $b^\prime_0=c=1$, $a_0^\prime=1$, so that the
rest diffusivity equals $D_0=1/2$.
Figure \ref{Fig_1} depicts a portion of a realization
of this process in the rest frame $\Sigma^\prime$,
and in an inertial frame moving with  respect to $\Sigma^\prime$
at constant velocity $w$.
Consider an ensemble of $N_p$ realizations of the Poisson-Kac
process. Each realization can be viewed as a stochastic particle,
the dynamics of which follows the microscopic dynamics 
eq. (\ref{eq1}). At time $t^\prime=t=0$ set $x^\prime=0$
for all the elements of the ensemble.
Figure \ref{Fig_2} depicts the overall probability density function $p(x,t)$
vs  the spatial coordinate $x$ of the moving inertial frame 
 obtained numerically from the stochastic simulations
over this ensemble for different values of $t$ and
for two different  relative velocity of the frame $\Sigma$: $w/c=0.8$ (panel a),
$w/c=0.95$ (panel b). All the simulations refer to a statistics over 
$N_p=10^5$  particles. 
As expected, the pdf's $p(x,t)$ are characterized by
a mean values that equals $-w \, t$, and by a variance $\sigma^2(t)$
that becomes narrower as $w$ increases up to the
limit value $c=1$ (compare the corresponding curves in the two panels).

\begin{figure}
\includegraphics[width=10cm]{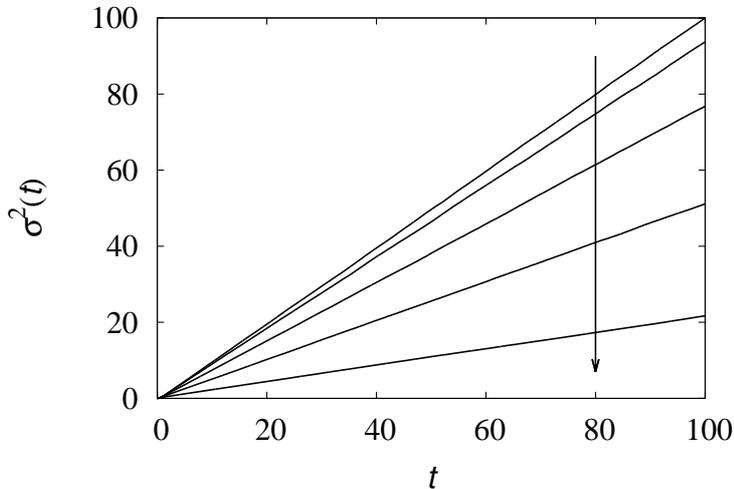}
\caption{$\sigma^2(t)$ vs $t$ for a 1d Poisson-Kac process ($b_0^\prime=c=1$, 
$a_0^\prime=1$)  in the  frame $\Sigma$ for different
values of the frame velocity $w$. The arrow indicates increasing values of
$w/c=0,\,0.2,\,0.4,\, 0.6,, 0.8$.}
\label{Fig_3}
\end{figure}

The graph of the variances $\sigma^2(t)$ vs $t$ for different
values of  the relative frame velocity $w$ is depicted
 in figure \ref{Fig_3}.
The arrow in this figure indicates increasing value of $w$ from $w=0$ (upper curve), up to $w=0.8$  (lower curve). For $w=0$, $D_0=1/2$ as expected from
eq. (\ref{eq3}), i.e. $\sigma^2(t)=t$, while as $w$ increases
the diffusion coefficient $D$, corresponding to half of  the slope
of the asymptotic linear plot of $\sigma^2(t)$  vs $t$, decreases.

Observe that the concept of long-term properties referred to
the recombination dynamics amongst the partial probability
waves $p^+$ and $p^-$ refers to time-scales $t \gg t_{\rm min}$,
where $t_{\rm min}=1/a^-$. For instance at $v=0.8$, $t_{\rm min}=3$,
corresponding to a very fast relaxation towards the asymptotic properties
compared to the time-scales reported in the abscissa of figure \ref{Fig_3}.

\begin{figure}
\includegraphics[width=12cm]{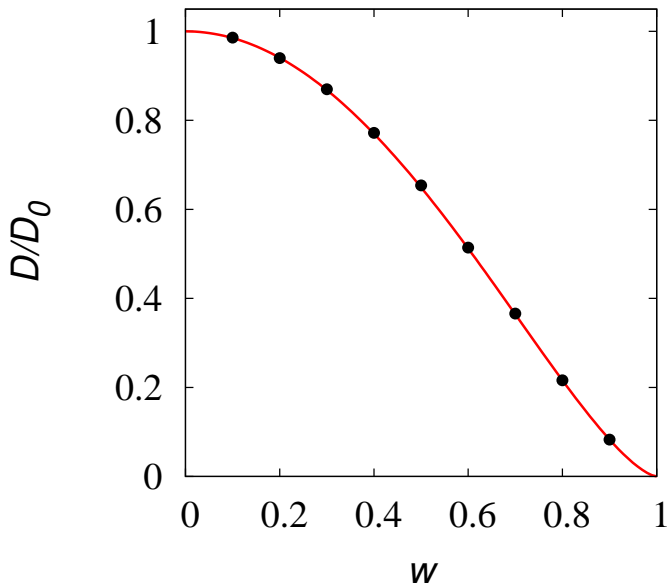}
\caption{$D/D_0$ vs $w$ ($c=1$) for a 1d Poisson-Kac process ($b_0^\prime=c=1$, $a_0^\prime=1$). 
Symbols ($\bullet$) represent the results of stochastic simulations,
solid line  the curve $D/D_0=\gamma^{-3}(w)=(1-w^2)^{3/2}$.}
\label{Fig_4}
\end{figure}

\begin{figure}[htb]
\begin{center}
\includegraphics[width=8.5cm]{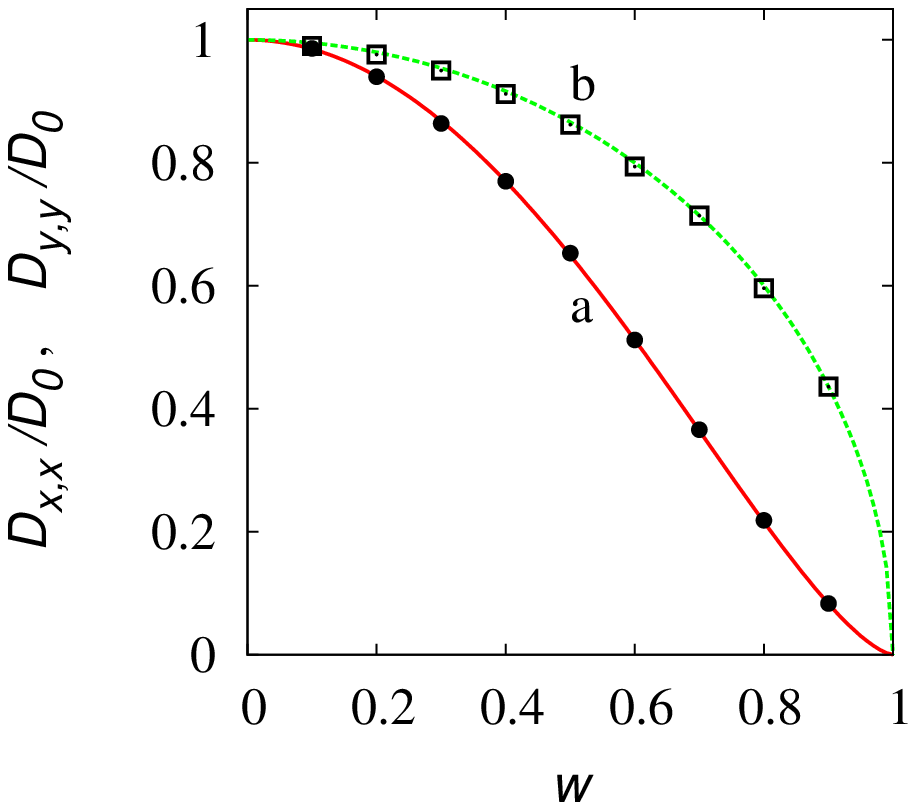}
\hspace{-2.2cm} {\Large (a)}
\includegraphics[width=8.5cm]{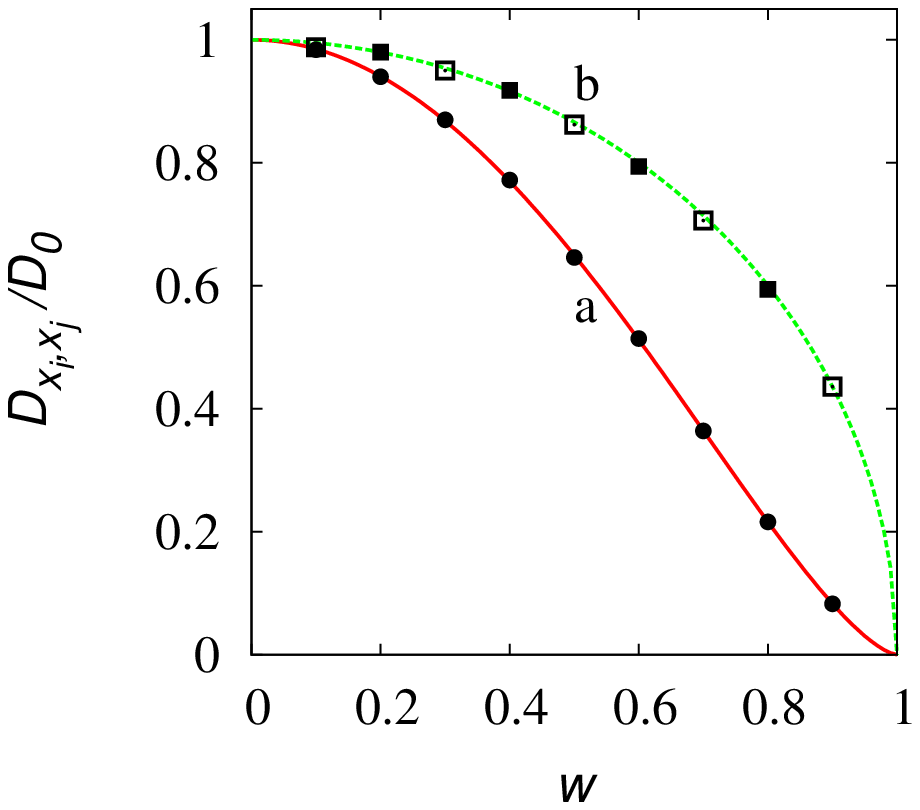}
\hspace{-2.2cm} {\Large (b)}
\end{center}
\caption{Panel (a): $D_{x,x}/D_0$ (symbols $\bullet$) and line (a)),
and $D_{y,y}/D_0$ (symbols $\square$) and line (b)) vs the relative frame
velocity $w$ for a two-dimensional Kolesnik-Kac process ($N=5$,
$b_0^\prime=c=1$, $a_0^\prime=1$).
Line (a) represents the function $\gamma^{-3}(w)$,
eq. (\ref{eq5_38}), line (b) the function $\gamma^{-1}(w)$,
eq. (\ref{eq5_44}). Symbols ($\bullet$) and ($\square$)
depict the results of stochastic simulations for $D_{x,x}$ and $D_{y,y}$,
respectively.
Panel (b): Diagonal diffusivities for the three-dimensional Poisson-Kac process
eq. (\ref{eq8_1}) with 
$b_{0^\prime}=c=1$, $a_0=1$. Symbols represent the results of stochastic
simulations: ($\bullet$) refer to $D_{x_1,x_1}$, ($\square$) and ($\blacksquare$) to $D_{x_2,x_2}$ and $D_{x_3,x_3}$, respectively. Lines (a) and (b) are
the same as in panel (a).}
\label{Fig_5}
\end{figure}

The behavior of the diffusion coefficient $D$ vs the  relative 
frame velocity $w$
is depicted in figure \ref{Fig_4} and it  is in perfect agreement
with  eq. (\ref{eq36}).
Next, consider higher dimensional stochastic processes.
The first example is given by the two-dimensional
Kolesnik-Kac model eqs. (\ref{eq5_1})-(\ref{eq5_2}).
We have chosen $N=5$, $b_0^\prime=c=1$,  $a_0^\prime=1$,
by considering an ensemble of $N_p=10^5$ stochastic particles.
Figure \ref{Fig_5} panel (a) shows the behavior of the
diagonal entries  $D_{x,x}$, $D_{y,y}$, obtained from
stochastic simulations  confirming the
theoretical expressions eqs. (\ref{eq5_38}) and (\ref{eq5_44}).
The off-diagonal  entry is not depicted  for the sake of brevity,
but is vanishing for any value of the frame velocity $w$.

Finally, let us consider a three-dimensional model, namely the
Poisson-Kac process
\begin{equation}
d x_i^\prime(t^\prime) = b_0^\prime \, (-1)^{\chi_i(t^\prime)} \, d t^\prime
\; , \qquad i=1,2,3
\label{eq8_1}
\end{equation}
where $\chi_i(t^\prime)$, $i=1,2,3$ are three Poisson processes,
statistically independent of each other, characterized by the
same values of the reference velocity $b_0^\prime$ and of the
transition rate $a_0^\prime$. This model, statistically described by means
of eight partial probability density functions, converges - for $b_0^\prime,
 a_0^\prime \rightarrow \infty$, keeping fixed the
 ratio $(b_0^\prime)^2/a_0^\prime$ -  to an
isotropic three-dimensional parabolic diffusion   equation for
the overall pdf $p({\bf x}^\prime,t^\prime)$ with
diffusion coefficient equal to $D_0$ given by eq. (\ref{eq3}).

Let $\Sigma$ be an inertial frame moving with a constant velocity
along the direction $x_1^\prime$ with respect to $\Sigma^\prime$. It is expected that 
eqs. (\ref{eq5_38}) and (\ref{eq5_44}) apply also in this
case, providing for the diffusivity tensor $D_{x_i,x_j}$ in $\Sigma$
the following expression
\begin{equation}
(D_{x_i,x_j})= D_0 \, \left (
\begin{array}{ccc}
\gamma^{-3}(v)  & \qquad  0 \qquad  & \qquad 0 \qquad \\
\qquad 0 \qquad & \gamma^{-1}(v)  & 0 \\
0 & 0 & \gamma^{-1}(v) 
\end{array}
\right ) 
\label{eq8_2}
\end{equation}
The results of stochastic simulations over an ensamble of $N_p=10^5$
particles are depicted in panel (b) of figure \ref{Fig_5} confirming
the theoretical prediction. Off-diagonal entries (not shown) prove to
be vanishing.

\section{Stochastic action invariance}
\label{sec8_bis}

There is a striking analogy between the results expressed
by eq. (\ref{eq8_2}) and the scaling of
the longitudinal and transveral masses.
Albeit the concepts of longitudinal and transversal masses,
introduced by Einstein  \cite{einstein_05} in the early days of relativity
theory (see also \cite{goldstein}), are nowadays
considered obsolete, their use in the
present context is convenient in order to derive
an interesting by-product of eq. (\ref{eq8_2}).

Consider the equation of motion of a particle of 
rest mass $m_0=E_0/c^2$ in an intertial frame
$\Sigma$ moving with velocity $w$ along the $x_1$-axis.
Introduce the diagonal mass tensor ${\bf m}=\mbox{diag}(m_\perp,m_\parallel, 
m_\parallel)$, where $m_\perp$ and $m_\parallel$ are the
longitudinal and transversal masses, respectively (the
mass tensor enters  the three-dimensional expression of
the relativistic Newton equation). 
The relativistic scaling of ${\bf m}$ is expressed by
\begin{equation}
{\bf m}= m_0 \,
\left (
\begin{array}{ccc}
\gamma^3(w) & 0 & 0 \\
0  & \gamma(w) & 0 \\
0 & 0 & \gamma(w) 
\end{array}
\right )
\label{eq8_bis_1}
\end{equation}
From eq. (\ref{eq8_2}) it follows that the
product ${\bf m} \, {\bf D}$ of a particle performing purely
diffusive stochastic 
motion is an isotropic tensor
\begin{equation}
{\bf m} \, {\bf D} = m_0 \, D_0 \, {\bf I}
\label{eq8_bis_2}
\end{equation}
and the non-vanishing diagonal entries are relativistically
invariant, i.e., they do not depend on the velocity $w$.

This observation, can be expressed as follows: if a particle
possessing rest mass $m_0$ evolves according to a relativistically
stochastic process admitting in a given inertial frame $\Sigma^\prime$
no convective contribution and an isotropic
diffusivity tensor ${\bf D}^\prime= D_0 \, {\bf I}$
(this specific inertial frame can be referred to as the
{\em rest frame} for the stochastic motion),
then in all the inertial frames $\Sigma$ moving
with respect to $\Sigma^\prime$ at constant velocity $w$, say along the
$x_1$-coordinate, the product ${\bf m} \, {\bf D}$ 
of the mass tensor times the diffusivity tensor 
is invariant with respect to $w$ and equal to $m_0 \, D_0 \, {\bf I}$.

The quantity $h_m=m_0 \, D_0$ has the physical dimension of 
$\mbox{kg} \, \mbox{m} ^2\mbox{/s}$,
i.e., it corresponds to an action, so that eq. (\ref{eq8_bis_3})
implies that the stochastic action $h_m$  is relativistically invariant.
It does not depend on the relative frame velocity, but eventually
it can depend on the particle rest mass.

We can develop further this concept by introducing quantum mechanical
considerations. From the well known correspondence between
the Schr\"{o}dinger equation and stochastic processes \cite{schr_stoca1,schr_stoca2}, usually obtained via an analytic continuation of the time coordinate
 towards the imaginary axis,
the quantum mechanical representation of the kinetic energy
of a free particle corresponds to an effective quantum diffusivity
$D_{\rm  quantum}$ expressed by
\begin{equation}
D_{\rm quantum}= \frac{\hbar}{2 \, m_0}
\label{eq8_bis_3}
\end{equation}
Therefore, for a free quantum particle, $D_0=D_{\rm quantum}$, and
eqs. (\ref{eq8_bis_2})-(\ref{eq8_bis_3})
provide
\begin{equation}
h_m = \frac{\hbar}{2}
\label{eq8_bis_4}
\end{equation}
i.e., the stochastic action is not only relativistically
invariant but also independent of the particle rest mass.
It indicates that in a stochastic representation of
quantum motion, the basic constraint induced by
quantization and by the  Lorentz transformation is the
relativistic invariance of the product of the mass
tensor times the effective diffusivity  tensor that,
in any inertial frame,  returns an isotropic tensor with
eigenvalues equal to $\hbar/2$,
\begin{equation}
{\bf m} \, {\bf D}= \frac{\hbar}{2} \, {\bf I}
\label{eq8_bis_5}
\end{equation}
Equation (\ref{eq8_bis_5}) can be viewed as a stochastic
quantization rule emerging from special relativity. Its implication
will be explored elsewhere. However, a qualitative indication
emerging from eq. (\ref{eq8_bis_5}) is that, even in the
low-energy limit, i.e., for the Schr\"{o}dinger equation (not speaking
of the Dirac counterpart),
a stochastic interpretation of its  formal structure
could be properly grounded on a relativistic covariant framework.
This implies that the source of stochasticity originating quantum
uncertainty, should also possess
covariant properties.  The natural candidate possessing
all these requirements is the zero-point energy fluctuations
of the electromagnetic field \cite{milonni}.
A thorough analysis of this issue will be addressed in
forthcoming works.

\section{Analysis of the diffusivity tensor: Space-Time Diffusion}
\label{sec9}

The results obtained for Poisson-Kac processes are confirmed
and generalized by the study of discrete stochastic space-time dynamics
which is addressed in this Section. This extension provides
a general expression for the relativitic transformation
of the diffusivity tensor.

\subsection{Space-Time Diffusion}
\label{sec9_1}

Discrete Space-Time Diffusion (henceforth STD) processes
in ${\mathbb R}^N$ has been originally introduced
in \cite{giona_std1} in order to describe in a compact way particle
transport in periodic arrays of obstacles or localized 
repulsive  potentials. A STD process is defined by
the quintuple $(N,S,\boldsymbol{\pi},
\{ {\bf A}_\alpha \}_{\alpha=1}^S,\boldsymbol{\tau})$,
where $N$ is the space-dimension, $S$ the number
of states the random space-time displacements can attain,
 $\boldsymbol{\pi}=(\pi_1,\dots,\pi_S)$  a $S$-dimensional
probability vector, $\pi_\alpha >0$, $\alpha=1,\dots,S$, $\sum_{\alpha=1}^S
\pi_\alpha=1$, ${\bf A}_\alpha$, $\alpha=1,\dots,S$,
 are given (constant) space-displacements
in ${\mathbb R}^N$, and $\boldsymbol{\tau}=(\tau_1,\dots,\tau_S)$
the corresponding time intervals $\tau_\alpha>0$.

Consider an inertial system $\Sigma$  defined by the space-time coordinates
$({\bf x},t)$.
The dynamics of a STD process in $\Sigma$ 
is defined with respect to a discrete
``iteration time'' $n=0,1,\dots$ as
\begin{equation}
({\bf x}_{n+1},t_{n})= ({\bf x}_n+ {\bf A}_\alpha,t_n+\tau_\alpha)
\qquad \mbox{Prob. } \pi_\alpha
\label{eq9_1}
\end{equation}
$\alpha=1,\dots,S$.
Suppose that the process is defined at $n=0$ such that ${\bf x}_0=0$,
$t_0=0$, so that no issues of simultaneity arise.
Eq. (\ref{eq9_1}) can be viewed as a stroboscopic sampling of
a stochastic process in the reference system $\Sigma$.

From the theory of STD processes developed in \cite{giona_std1},
the long-term evolution for the probability density $p({\bf x},t)$,
${\bf x}=(x_1,\dots,x_N)$,
associated with eq. (\ref{eq9_1}) converges to the solution
of an effective
constant-coefficient advection-diffusion equation
\begin{equation}
\partial_t p({\bf x},t) = -\sum_{k=1}^N v_k \, \partial_{x_k} p({\bf x},t)
+ \sum_{h,k=1}^N D_{h,k} \, \partial_{x_h} \partial_{x_k} p({\bf x},t)
\label{eq9_2}
\end{equation}
where $\partial_t=\partial /\partial t$, $\partial_{x_k}=\partial/\partial x_k$,
and ${\bf v}=(v_1,\dots,v_N)$, ${\bf D}=(D_{h,k})_{h,k=1}^N$
represent the constant effective velocity vector and tensor diffusivity,
respectively.

Introducing the quantities
\begin{eqnarray}
V_t^{(n)} & = & \sum_{\alpha=1}^S \pi_\alpha \, \tau_\alpha \, ,
\qquad V_k^{(n)} = \sum_{\alpha=1}^S \pi_\alpha \, A_{\alpha,k}
\,, \;\; k=1,\dots,N   \nonumber \\
D_t^{(n)} & = &   \frac{1}{2} \left [
\sum_{\alpha=1}^S \pi_\alpha \, \tau_\alpha^2 -
\left (V_t^{(n)} \right )^2 \right ] \, , \qquad 
\nonumber \\
 D_{t,k}^{(n)}  & =  &  \frac{1}{2} 
\left [\sum_{\alpha=1}^S \pi_\alpha \, \tau_\alpha 
\, A_{\alpha,k} - V_t^{(n)} \, V_k^{(n)} \right ] \,, \;\; k=1,\dots,N 
\nonumber \\
D_{h,k}^{(n)} & = &  \frac{1}{2} \left [
\sum_{\alpha=1}^S \pi_\alpha \, A_{\alpha,h}
\, A_{\alpha,k} - V_h^{(n)} \, V_k^{(n)} \right ] \,, \;\; h,k=1,\dots,N
\label{eq9_3}
\end{eqnarray}
where $A_{\alpha,k}$ is the $k$-th entry of the space-displacement
vector ${\bf A}_\alpha$,
the effective transport parameters attain the expression
\begin{equation}
v_k 
= \frac{V_k^{(n)}}{V_t^{(n)}} \, , \;\; \; k=1,\dots,N
\label{eq9_4}
\end{equation}
and
\begin {equation}
D_{h,k}=  \frac{ D_t^{(n)} \,V_h^{(n)} \, V_k^{(n)}}{\left ( V_t^{(n)} \right
)^3} - \frac{ \left [ D_{t,h}^{(n)} \, V_k^{(n)} + D_{t,k}^{(n)} \, V_h^{(n)}
\right ]}{\left ( V_t^{(n)} \right
)^2} + \frac{ D_{h,k}^{(n)}}{V_t^{(n)}} 
\label{eq9_5}
\end{equation}
$h=k=1,\dots,N$.
Observe that the quantities expressed by eqs. (\ref{eq9_3})
represent the effective space-time velocity and diffusivities
parametrized with respect to the discrete iteration time $n$.

\subsection{Relativistic analysis}
\label{sec9_2}

Let eq. (\ref{eq9_1}) be the dynamic description 
of a stochastic process in $\Sigma$, and let $\Sigma^\prime$ another inertial
frame, the space-time coordinate of which are $({\bf x}^\prime,t^\prime)$,
moving with respect to $\Sigma$ with constant relative velocity
$w$ along the $x_1$-axis.
Set $c=1$ for the light speed {\em in vacuo}, so
that $w \in (-1,1)$.

By the requirements of special relativity, the velocities
of the STD process (\ref{eq9_1}) should be bounded by $c$,
which implies that
\begin{equation}
|{\bf A}|_\alpha < \tau_\alpha \qquad  \alpha=1,\dots,S
\label{eq9_6}
\end{equation}

For simplicity, consider the case $N=2$, i.e., a spatial 
 two-dimensional 
model. In $\Sigma^\prime$ the STD process (\ref{eq9_1})
is described by the evolution equation
\begin{equation}
({\bf x}_{n+1}^\prime,t_{n}^\prime)= ({\bf x}_n^\prime+ {\bf A}_\alpha^\prime,t_n+\tau_\alpha^\prime)
\qquad \mbox{Prob. } \pi_\alpha
\label{eq9_7}
\end{equation}
$\alpha=1,\dots,S$, expressed with respect to the space-time coordinates
of $\Sigma^\prime$, where the displacements
${\bf A}_\alpha^\prime$, $\tau_\alpha^\prime$ are related to
${\bf A}_\alpha$, $\tau_\alpha$ by a Lorentz boost 
\begin{eqnarray}
\tau_\alpha^\prime & = & \gamma(w) \, (\tau_\alpha - w \, A_{\alpha,1})
\nonumber \\
A_{\alpha,1}^\prime & = & \gamma(w) \,  (A_{\alpha,1}  - w \, \tau_\alpha )
\label{eq9_8} \\
A_{\alpha,2}^\prime & = & A_{\alpha,2}
\nonumber
\end{eqnarray}
Given ${\bf A}_\alpha^\prime$ and $\tau_\alpha^\prime$, $\alpha=1,\dots,S$,
eqs. (\ref{eq9_3})-(\ref{eq9_5}) can be applied to
derive the effective transport parameters measured in $\Sigma^\prime$.

As regards the effective parameters with respect to the iteration
time $n$, elementary algebra provides
\begin{equation}
V_t^{\prime \, (n)} = \gamma(w) \, (V_t^{(n)} - w \, V_1^{(n)})
\, , \qquad V_1^{\prime \, (n)} = \gamma(w) \, (V_1^{(n)} - w \, V_t^{(n)})
\, , \qquad  V_2^{\prime \, (n)} = V_2^{(n)}
\label{eq9_9}
\end{equation}
and
\begin{eqnarray}
D_t^{\prime \, (n)} & = & \gamma^2(w) \, \left [ D_t^{(n)} - 2 \, w \,
D_{t,1}^{(n)} + w^2 \, D_{1,1}^{(n)} \right ]
\nonumber \\
D_{t,1}^{\prime \, (n)} & = & \gamma^2(w) \, \left [
(1+w^2) \, D_{t,1}^{(n)} - w \, \left (D_t^{(n)}+D_{1,1}^{(n)} \right ) 
\right ]  \nonumber \\
D_{t,2}^{\prime \, (n)} & = & \gamma(w) \, \left [ D_{t,2}^{(n)}
- w \, D_{1,2}^{(n)} \right ]
\label{eq9_10} \\
D_{1,1}^{\prime \, (n)} & = & \gamma^2(w) \, \left [
D_{1,1}^{(n)} - 2 \, w \, D_{t,1}^{(n)} + w^2 \, D_t^{(n)} \right ]
\nonumber \\
D_{1,2}^{\prime \, (n)} & = & \gamma(w) \, \left [ D_{1,2}^{(n)}
- w \, D_{t,2}^{(n)} \right ] \nonumber \\
D_{2,2}^{\prime \, (n)} & = & D_{2,2}^{(n)} \nonumber
\end{eqnarray}
For the effective velocities $v_k^\prime$ measured in
$\Sigma^\prime$, from eqs. (\ref{eq9_4}) and (\ref{eq9_9})
 one obtains 
\begin{equation}
v_1^\prime = \frac{V_1^{\prime \, (n)}}{V_t^{\prime \, (n)}}
= \frac{v_1 - w}{1- w \, v_1} \qquad
v_2^\prime = \frac{V_2^{\prime \, (n)}}{V_t^{\prime \, (n)}}
= \frac{v_2 \, \sqrt{1 - w^2}}{1-w \, v_1}
\label{eq9_11}
\end{equation}
that correspond to the velocity transformations
of  special relativity.

More interesting is the transformation of the effective tensor
diffusivity, i.e., how ${\bf D}^\prime$ measured in $\Sigma^\prime$
is related to ${\bf D}$.
To begin with, consider $D_{1,1}^\prime$. Eq. (\ref{eq9_5}),
expressed in $\Sigma^\prime$,
can be written in the form of an Euclidean scalar product $\langle \cdot,\cdot \rangle$, as
\begin{equation}
D_{1,1}^\prime = \frac{1}{(V_t^{\prime \, (n)})^3} \,
\langle \widetilde{\bf V}^\prime, \widetilde{\bf D}^\prime  \, 
\widetilde{\bf V}^\prime \rangle
\label{eq9_12}
\end{equation}
where
\begin{equation}
\widetilde{\bf V}^\prime = \left (
\begin{array}{c}
V_1^{\prime \, (n)} \\
V_t^{\prime \, (n)}
\end{array}
\right )
\, ,
\qquad
\widetilde{\bf D}^\prime = \left (
\begin{array}{cc}
D_t^{\prime \, (n)} & -D_{t,1}^{\prime \, (n)} \\
-D_{t,1}^{\prime \, (n)} & D_{1,1}^{\prime \, (n)}
\end{array}
\right )
\label{eq9_13}
\end{equation}
$\widetilde{\bf V}^\prime$ is related to $\widetilde{\bf V}=(V_1^{(n)},V_t^{(n)})$
by a Lorentz boost $\widetilde{\bf V}^\prime= \widehat{\mathcal L}_w \widetilde{V}$,
\begin{equation}
\widehat{\mathcal L}_w =
\left (
\begin{array}{cc}
\gamma & -\gamma \, w \\
-\gamma w & \gamma
\end{array}
\right )
\label{eq9_14}
\end{equation}
The transformation for the entries of $\widetilde{\bf D}^\prime$
stemming from eq. (\ref{eq9_10})
is compactly expressed by
\begin{equation}
\left (
\begin{array}{c}
D_t^{\prime \, (n)} \\
-D_{t,1}^{\prime \, (n)} \\
-D_{t,1}^{\prime \, (n)} \\
D_{1,1}^{\prime \, (n)} 
\end{array}
\right )
= \gamma^2(w) \,
\left (
\begin{array}{cccc}
1 & w & w & w^2 \\
w & \frac{1+w^2}{2} & \frac{1+w^2}{2} & w \\
w & \frac{1+w^2}{2} & \frac{1+w^2}{2} & w \\
w^2 & w & w & 1
\end{array}
\right )
\, 
\left (
\begin{array}{c}
D_t^{(n)} \\
-D_{t,1}^{(n)} \\
-D_{t,1}^{(n)} \\
D_{1,1}^{(n)} 
\end{array}
\right )
\label{eq9_16}
\end{equation}
The latter transformation can be expressed in tensorial form
as $\widetilde{D}_{h,k}^\prime=\Lambda_{h,k}^{m,n} \, \widetilde{D}_{m,n}$,
where the fourth-order tensor $\Lambda_{h,k}^{m,n}$ accounts for
the transformation (\ref{eq9_16}) and Einstein summation notation
has been adopted.
Consequently, eq. (\ref{eq9_12}) becomes
\begin{equation}
D_{1,1}^\prime = \frac{1}{\gamma^3(w) \, (1-w \, v_1)^3}
\, \frac{ M_{h,k}^{p,q} \, \widetilde{V}^h \, \widetilde{V}^k
\, \widetilde{D}_{p,q} }{\left ( V_t^{(n)} \right )^3}
\label{eq9_17}
\end{equation}
where $\widetilde{V}^h$, $\widetilde{D}_{p,q}$ are the
entries of $\widetilde{\bf V}$ and $\widetilde{\bf D}$, above
defined, and  $M_{h,k}^{p,q} = \widehat{\mathcal L}_{w,h}^p \, \Lambda_{p,q}^{h,k} \, \widehat{\mathcal L}_{w,k}^q$, $\widehat{\mathcal L}_{w,h}^p $ being the
entries of the Lorentz boost  (\ref{eq9_14}).
Developing the algebra, eq. (\ref{eq9_17}) yields the
following expression for $D_{1,1}^\prime$
\begin{equation}
D_{1,1}^\prime = \frac{1}{\gamma^3(w) \, (1-w \, v_1)^3}
= \left [
\frac{D_t^{(n)} \, (V_1^{(n)})^2 - 2 \, D_{t,1}^{(n)} \, V_1^{(n)}
\, V_t^{(n)} + D_{1,1}^{(n)} \, (V_t^{(n)})^2}{\left ( V_t^{(n)} \right )^3}
\right ]
\label{eq9_18}
\end{equation}
The term under square bracket in eq. (\ref{eq9_18}) is just
$D_{1,1}$ as measured in $\Sigma$, eq. (\ref{eq9_5}) for $h=k=1$.
Consequently, the trasformation for $D_{1,1}$ attains the
compact expression
\begin{equation}
D_{1,1}^\prime = \frac{D_{1,1}}{\gamma^3(w) \, (1-w \, v_1)^3}
\label{eq9_19}
\end{equation}
Next, consider $D_{1,2}^\prime$ that in $\Sigma^\prime$ is given by
\begin{equation}
D_{1,2}^\prime= \frac{D_t^{\prime \, (n)} \, V_1^{\prime \, (n)} \, V_2^{\prime \, (n)} 
- D_{t,1}^{\prime \, (n)} \, V_2^{\prime \, (n)} \, V_t^{\prime \, (n)}
- D_{t,2}^{\prime \, (n)} \, V_1^{\prime \, (n)} \, V_t^{\prime \, (n)}
+ D_{1,2}^{\prime \, (n)} \, (V_t^{\prime \, (n)})^2}{\left ( V_t^{\prime \, (n)}  \right )^3}
\label{eq9_20}
\end{equation}
Substituting the expressions (\ref{eq9_9})-(\ref{eq9_10})
 for the transport parameters in $\Sigma^\prime$
referred to the iteration time $n$ as a function of the
corresponding quantities in $\Sigma$, elementary algebra provides
the expression
\begin{equation}
D_{1,2}^\prime = \frac{D_{1,2}+ w \, Q_{1,2}}{\gamma^2(w) \, (1-w \, v_1)^3}
\label{eq9_21}
\end{equation}
where
\begin{equation}
Q_{1,2}= \frac{D_{1,1}^{(n)} \, V_2^{(n)} \, V_t^{(n)} - D_{1,2}^{(n)}
\, V_1^{(n)} \, V_t^{(n)}- D_{t_1}^{(n)} \, V_1^{(n)} \, V_2^{(n)}
+ D_{t,2}^{(n)} \, (V_1^{(n)})^2}{\left ( V_t^{\prime \, (n)} \right  )^3}
\label{eq9_22}
\end{equation}
Eq. (\ref{eq9_5}) can be used to express $D_{1,1}^{(n)}$ and $D_{1,2}^{(n)}$
as a function of $D_{1,1}$ and $D_{1,2}$. In this way,
the expression for $Q_{1,2}$ greatly simplifies, providing
$Q_{1,2}=D_{1,1} \, v_2 - D_{1,2} \, v_1$, and consequently
the tranformation relation for $D_{1,2}^\prime$ becomes
\begin{equation}
D_{1,2}^\prime = \frac{D_{1,2} + w \, (D_{1,1} \, v_2 - D_{1,2} \, v_1)}
{\gamma^2(w) \, (1- w \, v_1)^3}
\label{eq9_23}
\end{equation}
Finally, consider $D_{2,2}^\prime$. From its definition
\begin{equation}
D_{2,2}^\prime= \frac{D_t^{\prime \, (n)} \, ( V_2^{\prime \, (n)} )^2
-  2 \, D_{t,2}^{\prime \, (n)} \, V_2^{\prime \, (n)} \, V_t^{\prime \, (n)}
+ D_{2,2}^{\prime \, (n)} \, (V_t^{\prime \, (n)})^2}{\left ( V_t^{\prime \, (n)}  \right )^3}
\label{eq9_24}
\end{equation}
which can be rearranged in the form
\begin{equation}
D_{2,2}^\prime = \frac{D_{2,2} + 2 \, w \, P_{2,2} + w^2 \, R_{2,2}}{\gamma(w)
\, (1- w \, v_1)^3}
\label{eq9_25}
\end{equation}
where
\begin{eqnarray}
P_{2,2} & = & \frac{D_{1,2}^{(n)} \, V_2^{(n)} \, V_t^{(n)}
-D_{2,2}^{(n)} \, V_1^{(n)} \, V_t^{(n)}- D_{t,1}^{(n)} \, (V_2^{(n)})^2
+ D_{t,2}^{(n)} \, V_1^{(n)} \, V_2^{(n)}}{\left ( V_t^{\prime \, (n)}  \right )^3} \nonumber \\
R_{2,2} & = & \frac{D_{1,1}^{(n)} \, (V_2^{(n)})^2 - 2 \, D_{1,2}^{(n)}
\, V_1^{(n)} \, V_2^{(n)} + D_{2,2}^{(n)} \, (V_1^{(n)})^2}{\left ( V_t^{\prime \, (n)}  \right )^3}
\label{eq9_26}
\end{eqnarray}
Using eq. (\ref{eq9_5}) to express $D_{h,k}^{(n)}$ as
a function of the diffusivities $D_{h,k}$ expressed with
respect to the physical time, the expressions for
$P_{2,2}$ and $R_{2,2}$ simplifies to obtain
for $D_{2,2}^\prime$ the transformation
\begin{equation}
D_{2,2}^\prime= \frac{D_{2,2}+ 2 \, w (D_{1,2} \, v_2 - D_{2,2} \, v_1)
+ w^2 (D_{1,1} \, v_2^2 - 2 \, D_{1,2} \, v_1 \, v_2 + D_{2,2} \, v_1^2
)}{\gamma(w) \, (1-w \, v_1)^3}
\label{eq9_27}
\end{equation}

\subsection{The three-dimensional case}
\label{sec9_3}

The  extension of the transformations developed in the previous paragraph to
three-dimensional spatial coordinates in straightforward.
As before, suppose that $\Sigma^\prime$ moves relatively to $\Sigma$
with a constant velocity $w$ along the $x_1$-axis.

As  regards  $D_{1,1}^\prime$, $D_{1,h}^\prime=D_{h,1}^\prime$, $D_{h,h}^\prime$,
with $h=2,3$ the expressions follow from eqs.   (\ref{eq9_19}),
(\ref{eq9_23}) and (\ref{eq9_27}), namely
\begin{eqnarray}
D_{1,1}^\prime & =  & \frac{D_{1,1}}{\gamma^3(w) \, (1-w \, v_1)^3}
\nonumber \\
D_{1,h}^\prime & = & \frac{D_{1,h}+  w \, (D_{1,1} \, v_h - D_{1,h}
\, v_1)}{\gamma^2(w) \, (1-w \, v_1)^3} \qquad h=2,3
\label{eq9_28} \\
D_{h,h}^\prime & = & \frac{D_{h,h}  +  2 \, w \, (D_{1,h} \, v_h - D_{h,h}
\, v_1) + w^2 \, (D_{1,1} v_h^2 - 2 \, D_{1,h} \, v_1 \, v_h + D_{h,h}
\, v_1^2 ) }{\gamma(w) \,  (1-w \, v_1)^3}  \;\; \;\; h=2,3
\nonumber 
\end{eqnarray}
The entry $D_{2,3}^\prime=D_{3,2}^\prime$ requires some
additional calculations that, following the
same approach outlined in the previous paragraph,  returns
the expression
\begin{equation}
D_{2,3}^\prime = \frac{D_{2,3}+ w \, (D_{1,2} \, v_3 + D_{1,3} \, 
v_2 - 2 \, D_{2,3} \, v_1) + w^2 \, (D_{1,1}
\, v_2 \, v_3 - D_{1,2} \, v_1 \, v_3 - D_{1,2} \, v-1 \, v_3 + D_{2,3} \, v_1^2)}{\gamma(w) \, (1-w \, v_1)^3}
\label{eq9_29}
\end{equation}
This completes the analysis of the Lorentzian transformation
of the tensor diffusivity referred to two inertial frame in relative
motion.

\subsection{Numerical simulations}
\label{sec9_4}

In this paragraph, a numerical validation of the transformation theory
for the effective tensor diffusivities is provided.
Consider a spatially two-dimensional STD model, $N=2$ and  $S=3$, with
$\boldsymbol{\pi}=(0.5,0.1,0.3)$, $\boldsymbol{\tau}=(1,0.6,0.5)$,
and ${\bf A}_1=(0.8,0.4)$, ${\bf A}_2=(-0.4,0.3)$ and 
${\bf A}_3=(0.48,-0.05)$.

Numerical simulations of STD dynamics have been performed by considering
an ensemble of $10^7$ particles, initially located
at the same space-time point (${\bf x}_0$,$t_0=0$) evolving
according to eq. (\ref{eq9_1}). Using the Lorentz boost, the
corresponding coordinates ${\bf x}_n^\prime$, $t_n^\prime$
 in $\Sigma^\prime$ can be derived, and from the linear scaling of
the first and second-order (central) moments with  respect to 
$t^\prime$, the values 
$v_k^\prime$ and $D_{h,k}^\prime$ can be estimated.

\begin{figure}[htb]
\begin{center}
\includegraphics[width=10cm]{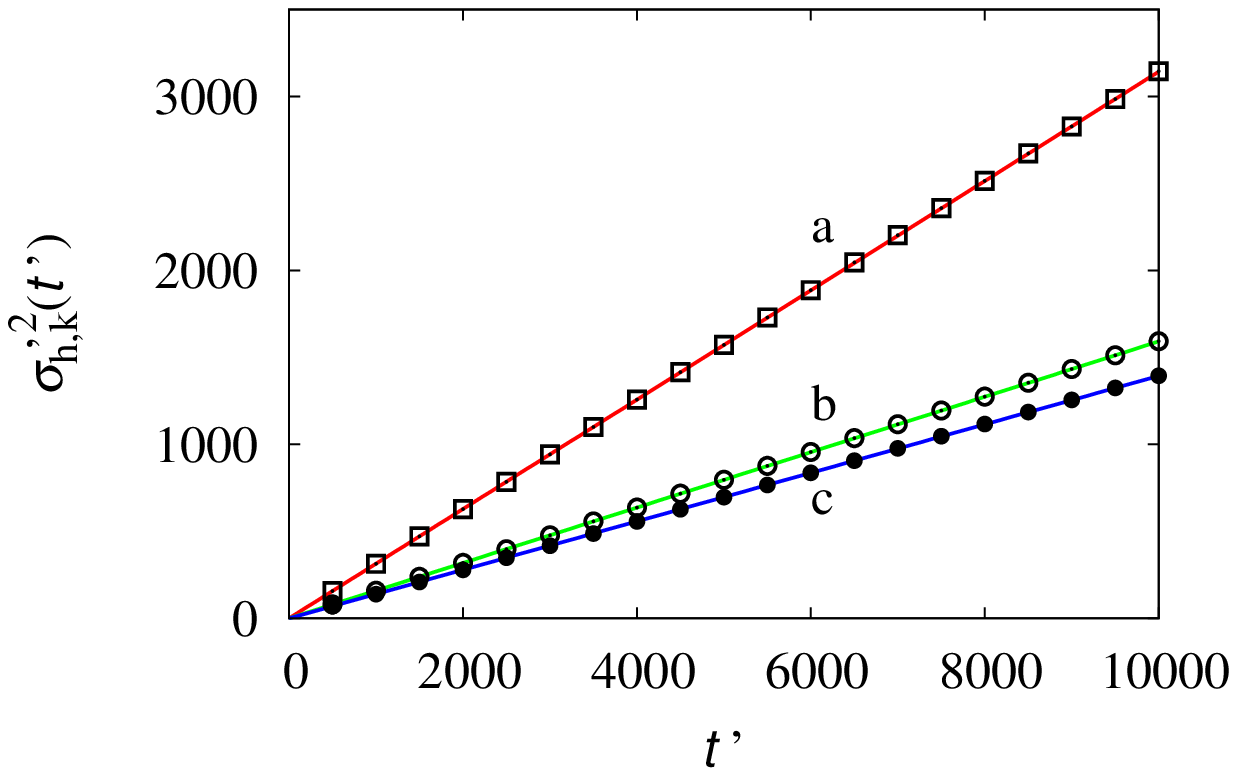}
\end{center}
\caption{Behavior of the second order central moments ${\sigma_{h,k}^{\prime \, 2}}(t^\prime)$ measured in $\Sigma^\prime$ vs $t^\prime$ for the two-dimensional spatial
model described in the main text. Symbols are the results of numerical
simulation of the stochastic dynamics, solid lines are the
theoretical predictions based on the Einstein scaling
${\sigma_{h,k}^{\prime \, 2}}(t^\prime) = 2 \, D_{h,k}^\prime \, t^\prime$,
where $D_{h,k}^\prime$ are given by eqs. (\ref{eq9_19}), (\ref{eq9_23}), (\ref{eq9_27}).
Line (a) and ($\square$) refer to ${\sigma_{1,1}^{\prime \, 2}}$,
line (b) and ($\circ$) to $\sigma_{1,2}^{\prime \, 2}$,
line (c) and ($\bullet$) to $\sigma_{2,2}^{\prime \, 2}$.}
\label{Fig_6}
\end{figure}

Figure \ref{Fig_6} depicts the scaling of the
second-order central moments $\sigma_{h,k}^{\prime \, 2}(t^\prime)=
\langle (x_h^\prime(t^\prime) - \langle x_h(t^\prime)\rangle) \,
(x_k^\prime(t^\prime) - \langle x_k(t^\prime)\rangle) \rangle$ for
$w=0.7$.
Solid lines represent  the theoretical predictions
$\sigma_{h,k}^{\prime, 2}(t^\prime)= 2 \, D_{h,k}^\prime \, t^\prime$,
where $D_{h,k}^\prime$ are given by eqs. (\ref{eq9_19}),
(\ref{eq9_23}) and (\ref{eq9_27}), while symbols refer
to the results of the numerical simulations of the stochastic
STD model.

\begin{figure}[htb]
\begin{center}
\includegraphics[width=10cm]{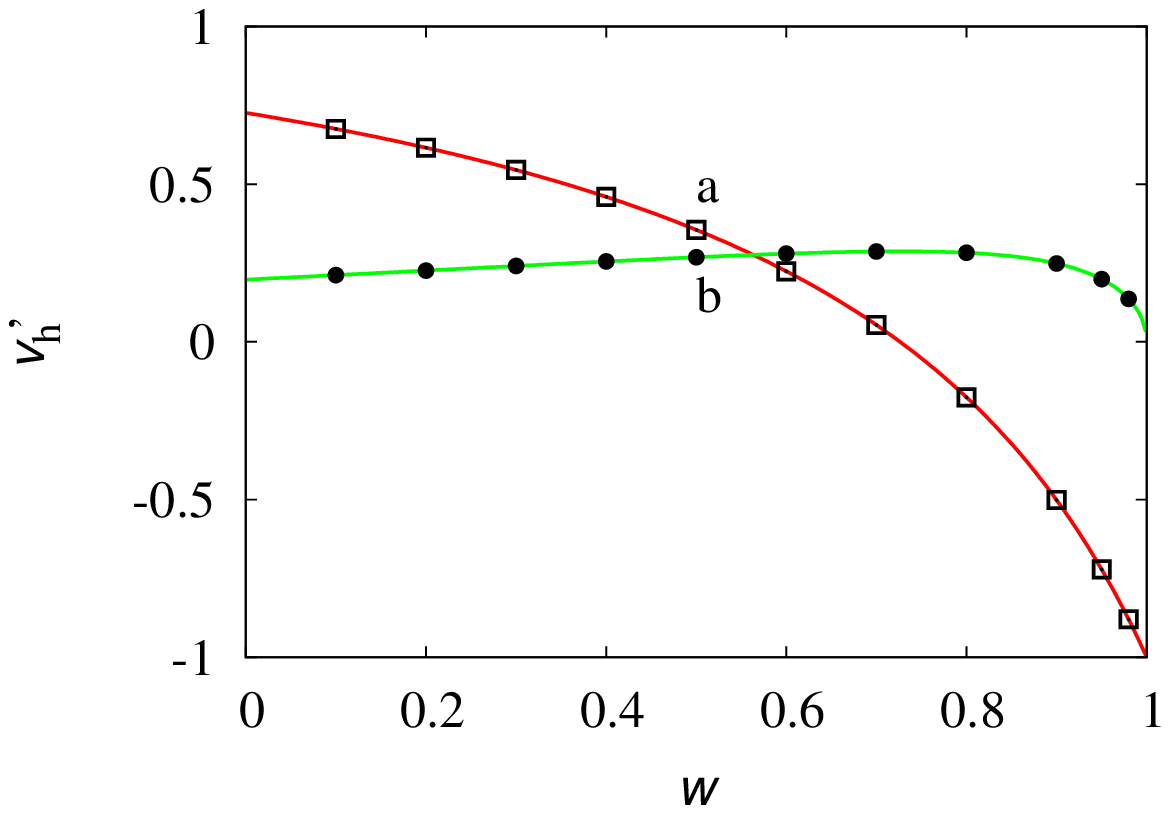}
\hspace{-1cm} {\Large (a)} \\
\includegraphics[width=10cm]{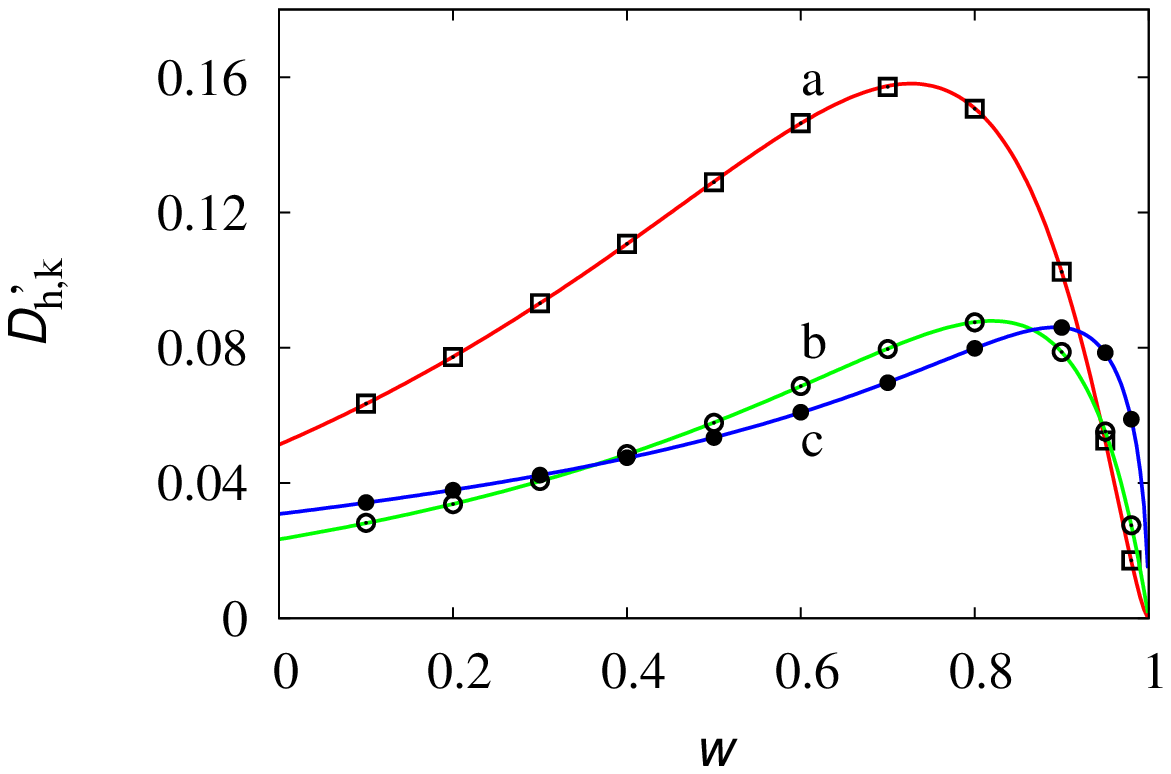}
\hspace{-1cm} {\Large (b)}
\end{center}
\caption{Effective transport parameters of the  spatial two-dimensional
 STD model described in the main text measured in $\Sigma^\prime$
as a function of the relative frame velocity $w$.
Panel (a): effective velocity entries. Line (a) and ($\square$)
refer to $v_1^\prime$, line (b) and ($\circ$) to $v_2^\prime$.
Panel (b): effective tensor diffusivities. Line (a) and ($\square$)
refer to $D_{1,1}^\prime$, line (b) and ($\circ$) to $D_{1,2}^\prime$,
line (c) and ($\bullet$) to $D_{2,2}^\prime$.}
\label{Fig_7}
\end{figure}

The review of the value of the effective transport
parameters  measured in $\Sigma^\prime$ vs the relative velocity
$w$ can be found in figure \ref {Fig_7}. Panel (a) refers
to the effective velocity entries, and panel (b) to
the effective tensor diffusivities.

Apart  from the excellent agreement between theory (lines) and simulations
(symbols),
the behavior of $D_{h,k}^\prime$ vs $w$ is highly non-trivial,
and the effective diffusivities display a local maximum at
some values $w_{h,k}^*$  of the relative velocity
that depend on $h$ and $k$. This
phenomenon is a consequence of the  presence
in the STD model considered of  
an  advective contribution, accounted for by the effective velocity 
 $v_1$, that is  significantly greater than zero. Consequently,
the factor $(1-w \, v_1)^3$ at the  denominator of the expressions
for $D_{h,k}^\prime$ modulates their behavior, determining
non monotonic effects vs $w$. For example,
in the case of $D_{1,1}^\prime$, the abscissa $w_{1,1}^*$ of the
local maximum equal $v_1$ itself, in the present case $v_1 \simeq 0.726$,
and $D_{1,1}^\prime(w_{1,1}^*)$ is almost three times larger than
$D_{1,1}$.

For $w \rightarrow 1$, all the diffusivities $D_{h,k}^\prime$
decay to zero, as $D_{1,1}^\prime \sim \gamma^{-3}(w)$,
 $D_{2,2}^\prime \sim \gamma^{-1}(w)$,  $D_{1,2}^\prime \sim \gamma^{-2}(w)$.

\begin{figure}[htb]
\begin{center}
\includegraphics[width=10cm]{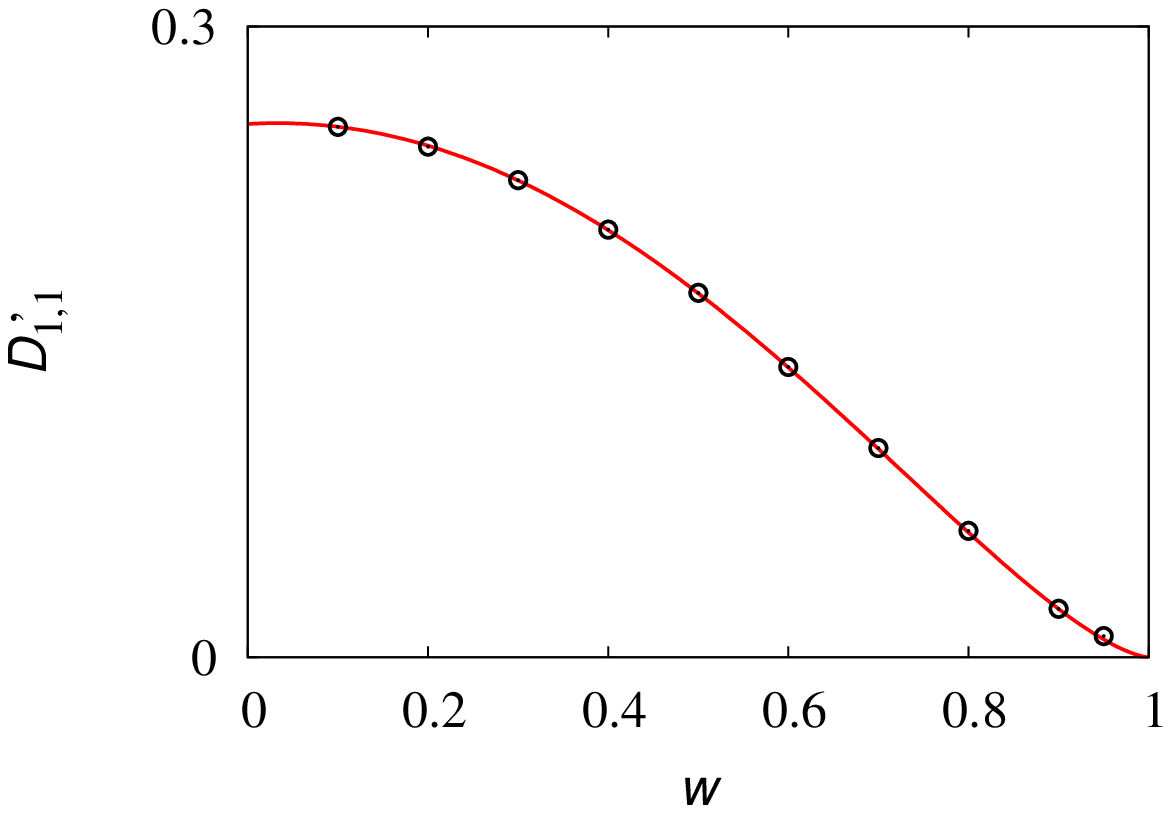}
\hspace{-1cm} {\Large (a)} \\
\includegraphics[width=10cm]{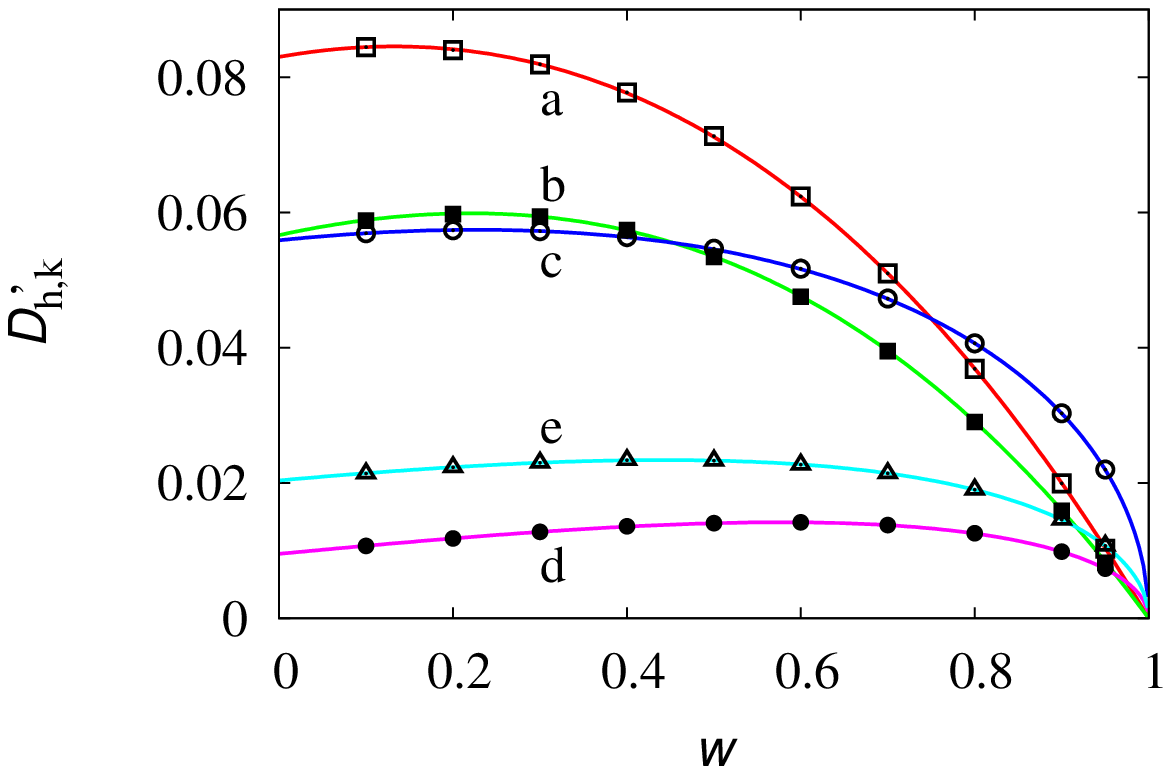}
\hspace{-1cm} {\Large (b)}
\end{center}
\caption{Effective tensor diffusivities $D_{h,k}^\prime$ vs the
relative frame velocity $w$ for the  spatial three-dimensional
model described in the main text. Panel (a): $D_{1,1}^\prime$
vs $w$. Panel (b): $D_{h,k}^\prime$ vs $w$.
Line (a) and ($\square$) refer to $D_{1,2}^\prime$, line
(b) and ($\blacksquare$) to $D_{1,3}^\prime$, line (c) and ($\circ$)
to $D_{2,2}^\prime$, line (d) and ($\bullet$) to $D_{2,3}^\prime$,
line (e) and ($\triangle$) to $D_{3,3}^\prime$.}
\label{Fig_8}
\end{figure}

The behavior of the tensor diffusivities in $\Sigma^\prime$ for
a spatially three-dimensional STD model ($N=3$) is depicted in figure
\ref{Fig_8}.
The STD model considered admits $S=6$ states with
$\boldsymbol{\pi}=(0,1,0.2,0.3,0.1,0.2,0.1)$,
$\boldsymbol{\tau}=(1,2,0.5,5,3,0.8)$ and
${\bf A}_1=(0.5,-0.3,0.2)$, ${\bf A}_2=(1,-0.4,0.2)$, 
${\bf A}_3=(1,0.05,-0.05)$, ${\bf A}_4=(-1.5,0.4,0.2)$,
${\bf A}_5=(1.5,1,0.6)$, ${\bf A}_6=(0.3,-0.2,0.3)$. Also
in this case, the agreement of the theoretical predictions (solid lines)
based on 
eqs. (\ref{eq9_28})-(\ref{eq9_29}) with respect to
the stochastic simulation data (symbols)
is excellent.

\section{Discussion and implications}
\label{sec10}

In this Section, some implications and observations related to
 the transformation
theory of tensor diffusivities are  addressed.

\subsection{Observation I - Generality of the transformation}
\label{sec10_1}

Although we have considered specific stochastic
kinematics - Poisson-Kac processes and STD dynamics -
eqs.  (\ref{eq9_28})-(\ref{eq9_29}) are of general validity.
To support and confirm this claim let us consider
a totally different problem, fairly unusual in a
relativistic context.

Consider a classical Langevin equation in ${\mathbb R}^3$
\begin{equation}
d x_h(t) = v_h \, dt + \sqrt{2 \, D_h} \, d w_h(t) \,,
\qquad h=1,2,3
\label{eq10_1}
\end{equation}
where $w_h(t)$, $h=1,2,3$ are the realizations of three
independent Wiener processes, and $v_h$, $D_h$ are constant.
This model is obviously relativistically inconsistent as,
by definition, a Wiener process possesses infinite propagation
velocity (and so do the  $x_h(t)$ defined by eq. (\ref{eq10_1})),
due to its Gaussian distribution of increments. Therefore,
in order to use eq. (\ref{eq10_1}) in the present analysis,
this process should be somehow ``cured''. The
``cure'' we apply is conceptually similar to the classical
Wong-Zakai mollification of Brownian motion \cite{wongzakai1,wongzakai2},
see also \cite{wzt}.

Let $\{ \widetilde{\bf x}_\alpha(t) \}_{\alpha=1}^{N_p}$
be an ensemble of $N_p$ particles moving according to
the stochastic kinematics (\ref{eq10_1}), starting from
${\bf x}_\alpha(t=0)=0$, and let $\{\widetilde{\bf x}_\alpha^{(n)}
= {\bf x}_\alpha(nT)\}_{\alpha=1}^{N_p}$, $n=0,1,\dots$ be their stroboscopic sampling
at multiples of the time interval $T>0$. In order to
make the sampling $\{\widetilde{\bf x}_\alpha^{(n)}\}_{\alpha=1}^{N_p}$
of the stochastic dynamics relativistically consistent, particle
positions should be filtered in order to ensure a propagation velocity
less than  $c$ ($c=1$ in the present analysis).
Let $\{{\bf x}_\alpha^{(n)}\}_{\alpha=1}^{N_p}$
be the filtered stroboscopic sampling of 
$\{\widetilde{\bf x}_\alpha^{(n)}\}_{\alpha=1}^{N_p}$, obtained in the
following way: (i) ${\bf x}_\alpha^{(0)}=\widetilde{\bf x}_\alpha^{(0)}=0$,
for all $\alpha=1,\dots,N$;  (ii) choose a reference
maximum velocity $v_{\rm max}<c$, say $v_{\rm max}=0.99$;
(iii) if $ \delta_\alpha^{(n)}=|| \widetilde{\bf x}_\alpha^{(n)}-{\bf x}_\alpha^{(n-1)} ||
> v_{\rm max} T$, i.e., if the relativistic velocity constraint could be
 locally
violated, set for ${\bf x}_\alpha^{(n)}$ the value
\begin{equation}
{\bf x}_\alpha^{(n)}= \frac{v_{\rm max} \, T}{\delta_\alpha^{(n)}}
\, \widetilde{\bf x}_\alpha^{(n)} + \left (
1- \frac{v_{\rm max} \, T}{\delta_\alpha^{(n)}} \right ) \,
{\bf x}_\alpha^{(n-1)}
\label{eq10_2}
\end{equation}
corresponding to a local propagation velocity equal to $v_{\rm max}$.
By definition, the filtered stroboscopic sampling
 $\{{\bf x}_\alpha^{(n)}\}_{\alpha=1}^{N_p}$, $n=0,1,2,\dots$ 
is relativistically consistent and
can be viewed as a form of Wong-Zakai mollification of the
original process, by considering the stochastic trajectories
between time instants $(n-1) T$, and $n T$ represented by
straight lines connecting
 ${\bf x}_\alpha^{(n-1)}$ to ${\bf x}_\alpha^{(n)}$.

Consequently, $\{ {\bf x}_\alpha^{(n)}\}_{\alpha=1}^{N_p}$ can be
viewed as an ensemble of realizations of a  relativistically
plausible stochastic
process defined in an inertial system $\Sigma$, out of which
its transport parameters can be estimated. If the velocities
and the diffusivities  are small enough, the effective transport
parameters estimated in $\Sigma$ practically
coincide with $v_h$ and $D_h$, i.e., with the
velocity and diagonal entries of the diffusivity
tensor entering eq. (\ref{eq10_1}).
Enforcing the Lorentz boost, the effective transport parameters
estimated in a reference $\Sigma^\prime$ moving with
respect to $\Sigma$ with constant relative velocity
$w<1$ along the $x_1$-axis can be obtained.

\begin{figure}[htb]
\begin{center}
\hspace{-1.0cm}
\includegraphics[width=6.5cm]{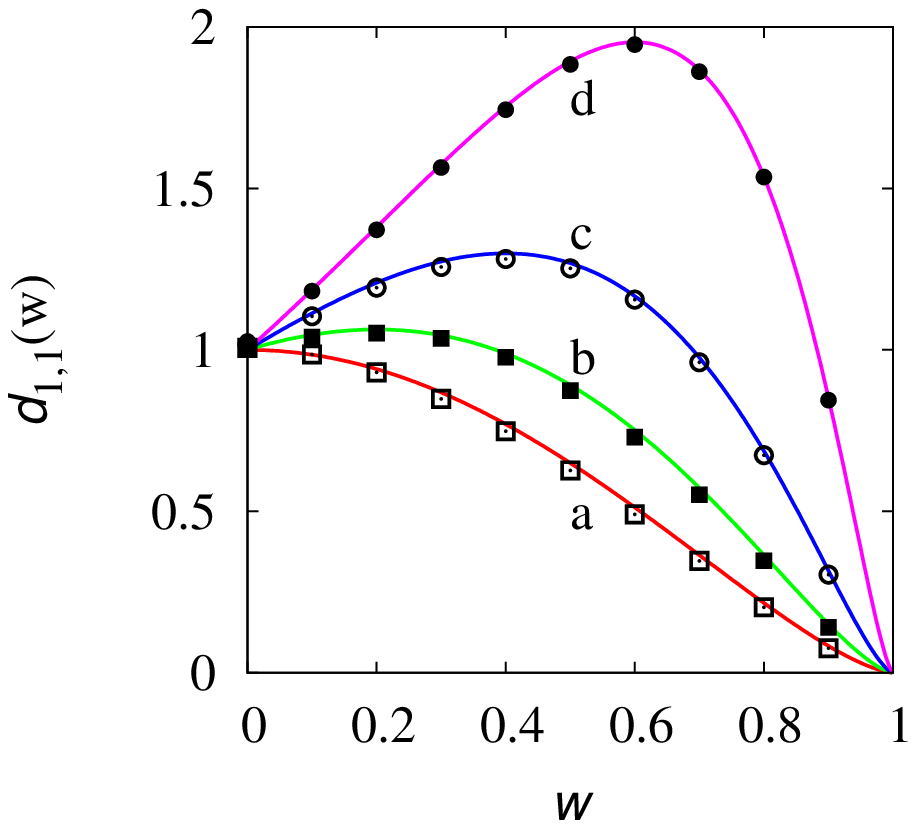}
\hspace{-1.cm} {\Large (a)}
\includegraphics[width=6.5cm]{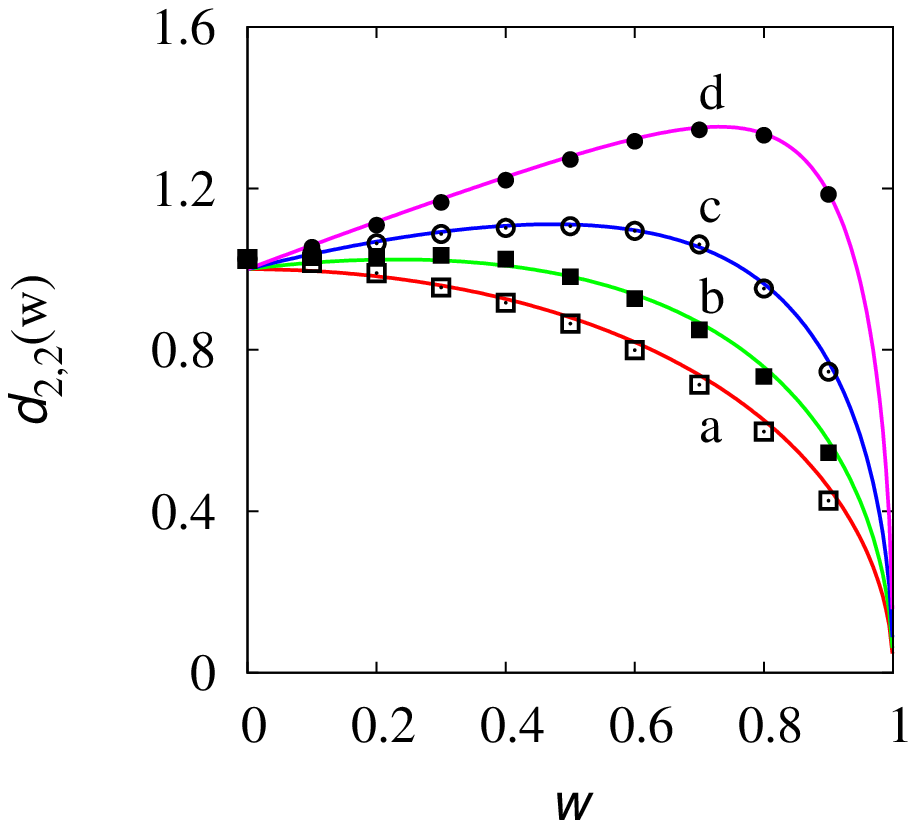}
\hspace{-1.cm} {\Large (b)} \\
\hspace{-1.0cm}
\includegraphics[width=6.5cm]{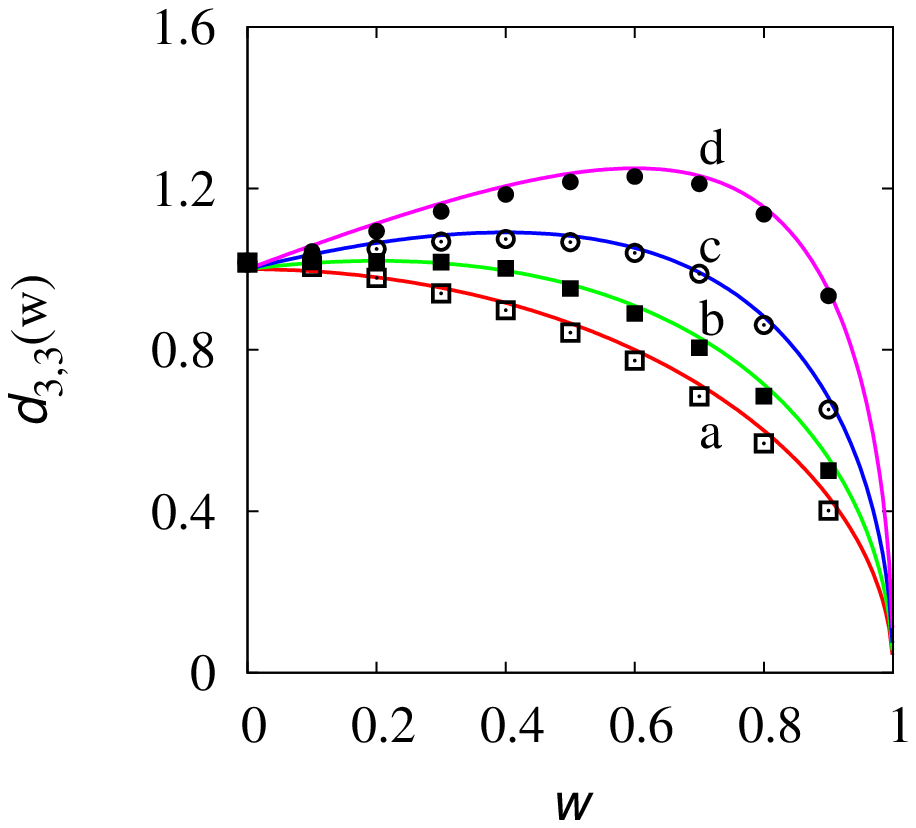}
\hspace{-1.cm} {\Large (c)} 
\includegraphics[width=6.5cm]{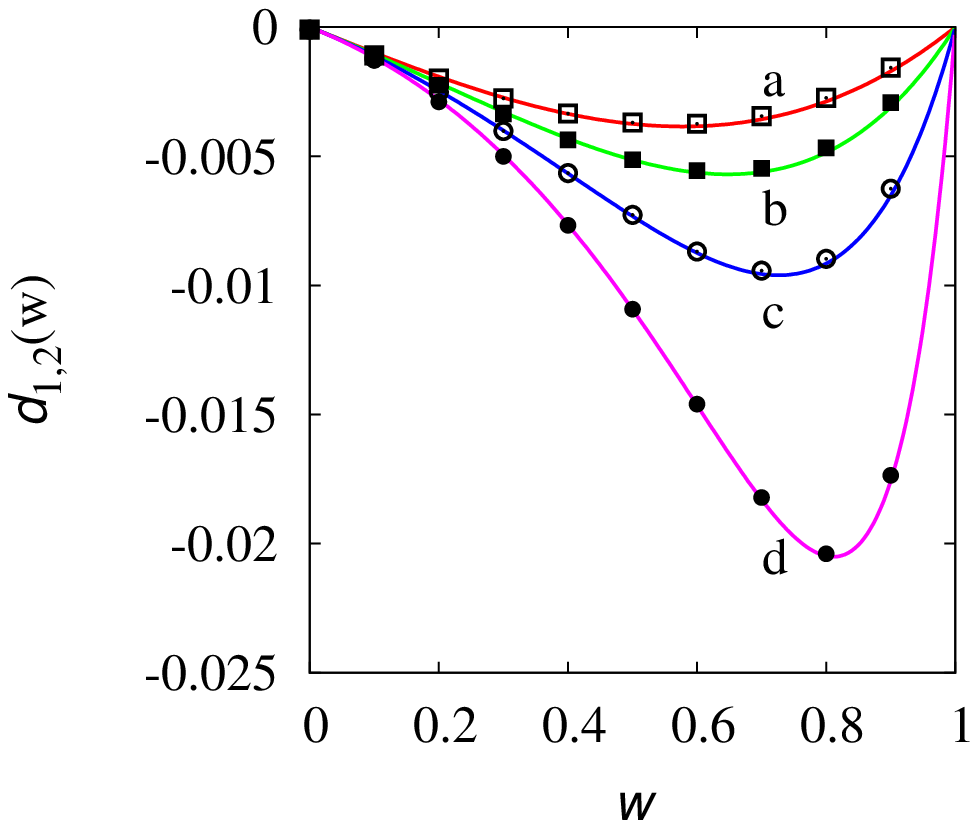}
\hspace{-1.cm} {\Large (d)} \\
\hspace{-1.0cm}
\includegraphics[width=6.5cm]{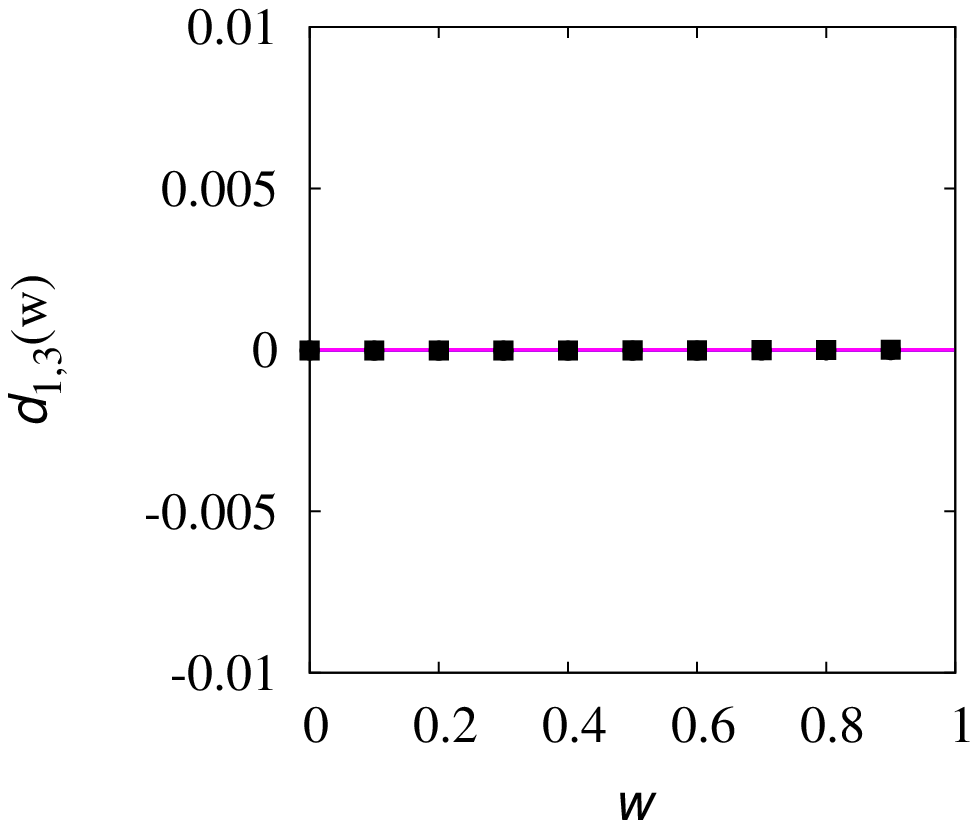}
\hspace{-1.cm} {\Large (e)}
\includegraphics[width=6.5cm]{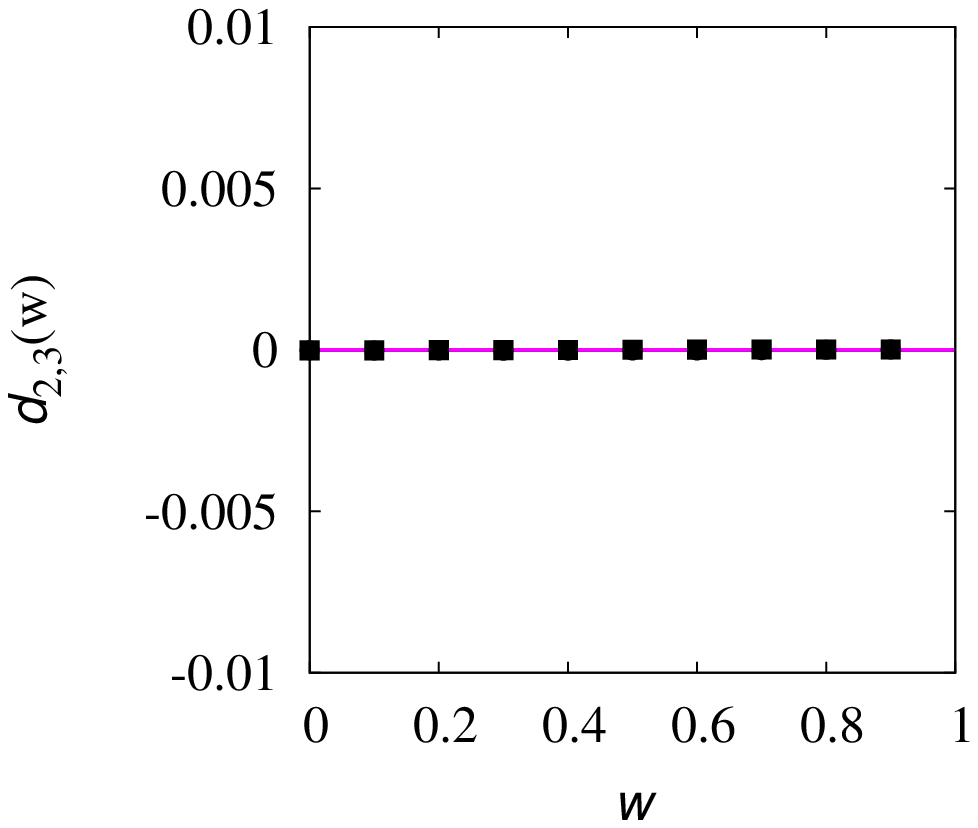}
\hspace{-1.cm} {\Large (f)}
\end{center}
\caption{$d_{h,k}(w)$ vs $w$ for the filtered Wong-Zakai mollification
of the process (\ref{eq10_1}) sampled at $T=20$.
Lines refer to eqs. (\ref{eq9_28})-(\ref{eq9_29}), symbols
to numerical simulations: lines (a) $a=0$, (b) $a=0.2$,
(c) $a=0.4$, (d) $a=0.6$. Panel (a) refers to $d_{1,1}(w)$,
panel (b) to $d_{2,2}(w)$, panel (c) to $d_{3,3}(w)$,
panel (d) to $d_{1,2}(w)=d_{2,1}(w)$, panel (e) to $d_{1,3}(w)=d_{3,1}(w)$,
panel (f) to $d_{2,3}(w)=d_{3,2}(w)$.}
\label{Fig_9}
\end{figure}

Consider for the  velocities and diffusivities entering
eq. (\ref{eq10_2}) the values
$v_1=a$, $v_2=-0.2$, $v_3=0$, $D_1=0.05$, $D_2=0.03$, $D_3=0.01$,
where $a$ is a parameter, and let $d_{h,h}(w)=D_{h,h}^\prime(w)/D_{h}$,
for $h=1,2,3$, and $d_{h,k}(w)=D_{h,h}^\prime(w)$ for $h \neq k$.
Figure \ref{Fig_9} panels (a) to (f) depict the six independent
entries of $d_{h,k}(w)$ as a function of the frame velocity $w$ at four 
different values of the parameter $a=0,\, 0.2,\, 0.4,\, 0.6$ controlling
 convective  particle motion along the $x_1$-direction.

Simulation results refer to the ensamble $\{ {\bf x}_\alpha^{(n)} \}_{\alpha=1}^{N_p}$ obtained from eq. (\ref{eq10_1}) using the Wong-Zakai filtering
discussed above with a sampling time $T=20$, and $N_p=10^5$.
Stochastic trajectories have been obtained from eq. (\ref{eq10_1})
using a classical Euler-Langevin algorithm with a time step $\Delta t=10^{-3}$.
In the present simulations, involving a rather small ensemble
of particles, eq. (\ref{eq10_2}) has never been used,
and solely the Wong-Zakai linear interpolation between
${\bf x}_\alpha^{(n-1)}$ and ${\bf x}_\alpha^{(n)}$ has been
applied in order to
obtain the particle position at constant value of time
$t^\prime$ measured in $\Sigma^\prime$.
Solid lines in figure \ref{Fig_9} refer to the theoretical
predictions based on eqs. (\ref{eq9_28})-(\ref{eq9_29}),
where $D_{h,k}=D_h \, d_{h,k}$, and $D_h$  are the
diagonal diffusivity entering eq. (\ref{eq10_1}). An excellent
agreement between theory and stochastic simulations can be
observed, confirming the general validity
of eqs. (\ref{eq9_28})-(\ref{eq9_29}).

\subsection{Observation 2 - Poisson-Kac processes and the limit
for $w \rightarrow c$}
\label{sec10_2}

The analysis developed for STD processes is in agreement with
the results obtained for the relativistic kinematics of Poisson-Kac
processes. In the latter case $D_{1,2}=0$, and the convective
contributions are absent, i.e. $v_1=v_2=0$. Correspondingly,
eqs. (\ref{eq9_28})-(\ref{eq9_29}) predict the
scaling of the longitudinal $D_{1,1}^\prime$ and transversal
$D_{2,2}^\prime$ diffusivities given by eq. (\ref{eq8_2}).

Particularly interesting is the limit of these equations
for $w$ approaching $c=1$. In this case the measured diffusivities in
$\Sigma^\prime$ vanish identically. For an observer
that moves close to  the velocity of light the contribution
of external stochastic perturbations, the intensity of which are
related to the diffusivities $D_{h,k}^\prime$, becomes
progressively irrelevant as $w \rightarrow c$.
This issue is discussed in greater detail in \cite{giona0}
which addresses the relativistic relevance of Poisson-Kac processes
as the prototype of a covariant stochastic kinematics.

In point of fact, the vanishing properties
 of $D_{h,k}^\prime$ for $w \rightarrow c$
could have a deeper physical meaning: it indicates that in a reference
system moving with a velocity approaching that of light all
the external dissipative and irreversible processes associated with
stochastic fluctuations decay to zero. 

\subsection{Observation 3 - Relativity of stochasticity and determinism}
\label{sec10_3}

There is another interesting implication of the transformation
theory of the tensor diffusivity.
From eqs. (\ref{eq9_28})-(\ref{eq9_29}) it follows
that the effective  diffusivities $D_{h,k}^\prime$ measured
in $\Sigma^\prime$ depends on convective velocities $v_h$.
Apart from the term $(1-w v_1)^{-3}$, this dependence enters
as factors multiplying the ${\mathcal O}(w)$, and ${\mathcal O}(w^2)$
terms in the expressions for $D_{h,k}^\prime$. The
only  diffusivity entry that is not influenced by
these convective contributions is the longitudinal
diffusivity  since $D_{1,1}^\prime=D_{1,1}/\gamma^3(w) (1-w v_1)^3$.

This observation suggests that it may happen that a process
that is regarded as fully deterministic in a reference system
$\Sigma$ appears to possess a stochastic nature in $\Sigma^\prime$
and viceversa. To clarify this concept it is convenient
to consider a simple example. Consider a STD process in $\Sigma$
(model I)
with $N=2$, $S=2$ characterized by the following parameters
\begin{equation}
\pi_1=\pi_2= \frac{1}{2} \, , \qquad
\tau_1=\tau_2= 1 \, , \qquad
{\bf A}_{1} =
\left (
\begin{array}{c}
0.8 \\
0.5 
\end{array}
\right )
\, ,
\;\;\;
{\bf A}_{2} =
\left (
\begin{array}{c}
-0.8 \\
0.5 
\end{array}
\right )
\label{eq10_3}
\end{equation}
For this process $v_1=0$, $v_2=0.5$, $D_{1,1}=0.32$,
$D_{1,2}=D_{2,1}=D_{2,2}=0$. Let $x=x_1$, $y=x_2$.
 In the long-term  limit, the dynamic of this process in $\Sigma$ is described
by the probability density function $p(x,y,t)$ that approaches the
solution of the parabolic transport equation
\begin{equation}
\partial_t p(x,y,t) = -v_2 \, \partial_y p(x,y,t) + D_{1,1} \,
\partial_x^2 p(x,y,t)
\label{eq10_4}
\end{equation}
The marginal probability density $p_y(y,t)=\int_{-\infty}^\infty
p(x,y,t) \, d x$ of the $y$-process satisfies a
strictly deterministic advection equation
\begin{equation}
\partial_y p_y(y,t) = - v_2 \, \partial_y p_y(y,t)
\label{eq10_5}
\end{equation}
which follows directly form the inspection of the
structure of the space-time displacements (\ref{eq10_3})
characterizing this model.
Viewed from $\Sigma$, the evolution of the $y$-dynamics defines
a strictly deterministic process.

\begin{figure}[htb]
\begin{center}
\includegraphics[width=10cm]{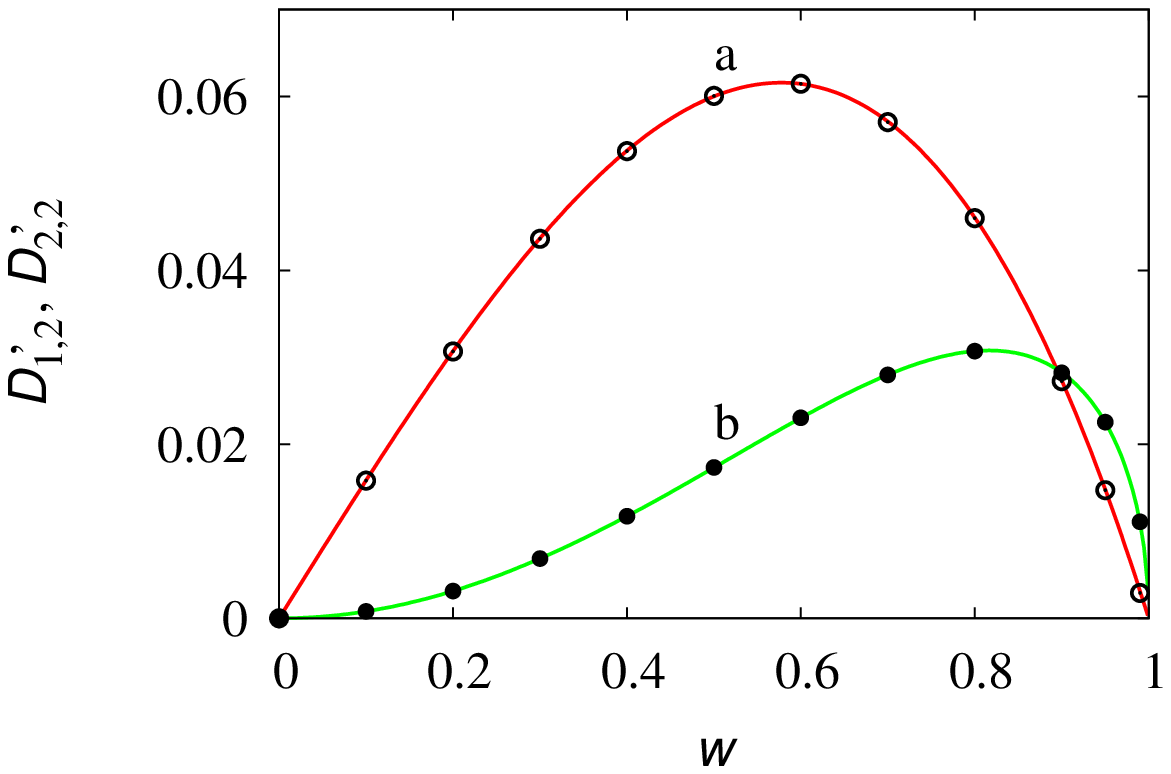}
\end{center}
\caption{Diffusivities $D_{1,2}^\prime$ and $D_{2,2}^\prime$
measured in $\Sigma^\prime$
vs the relative frame velocity $w$ for model I discussed in the main
text. Solid lines represent the theoretical predictions (\ref{eq9_28})-(\ref{eq9_29}),  symbols
the results of the numerical simulations of the stochastic STD dynamics
(\ref{eq10_3}).
Line (a) and ($\circ$) refer to $D_{1,2}^\prime$, line (b) and ($\bullet$)
to $D_{2,2}^\prime$.}
\label{Fig_10}
\end{figure}

Consider the same process viewed by $\Sigma^\prime$ moving with
respect to $\Sigma$ with constant relative velocity $w>0$ along
the $x$ axis. In this case the entries $D_{1,2}^\prime$ and $D_{2,2}^\prime$
of the diffusivity tensor are different from zero, and specifically
\begin{equation}
D_{2,2}^\prime = \frac{D_{1,1} \, w^2 v_2^2}{\gamma(w) \, (1-w \, v_1)^3}
>0
\label{eq10_6}
\end{equation}
This phenomenon is depicted in figure \ref{Fig_10} where the
theoretical expressions for $D_{1,2}^\prime$ and $D_{2,2}^\prime$
(solid lines) are compared with numerical simulations
of the STD (\ref{eq10_3}).

Therefore, the marginal probability density function $p^\prime_{y^\prime}(y^\prime, t^\prime)$
for the transversal $y^\prime$-process in $\Sigma^\prime$
approaches the solution of the advection-diffusion
equation
\begin{equation}
\partial_{t^\prime} p_{y^\prime}^\prime(y^\prime,t^\prime)=
- v_2^\prime \, p_{y^\prime} + D_{2,2}^\prime \, \partial_{y^\prime}^2 p_{y^\prime}^\prime(y^\prime,t^\prime)
\label{eq10_7}
\end{equation}
corresponding to the evolution of a stochastic process characterized
by an effective diffusivity $D_{2,2}^\prime >0$.
\begin{figure}[htb]
\begin{center}
\includegraphics[width=10cm]{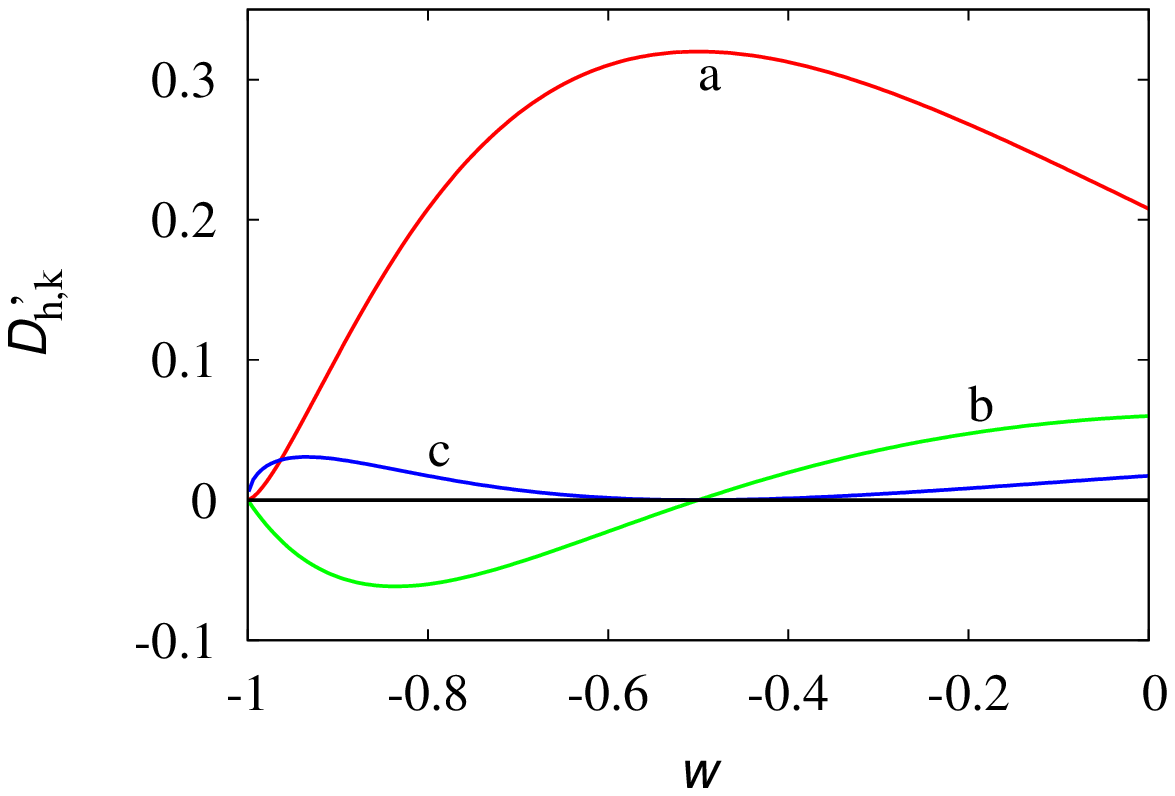}
\end{center}
\caption{Diffusivities $D_{h,k}^\prime$ vs the relative frame velocity $w$
measured in $\Sigma^\prime$
for model II discussed in the main text. Line (a) refers to $D_{1,1}^\prime$,
line (b) to $D_{1,2}^\prime$, line (c) to $D_{2,2}^\prime$.}
\label{Fig_11}
\end{figure}

The reverse is also true, by adopting the same argument.
Consider a stochastic process in $\Sigma$ for which
$D_{2,2} \neq 0$. By tuning the relative velocity $w$ of
$\Sigma^\prime$, it can happen that $D_{2,2}^\prime=0$.
Consequently, what appears in $\Sigma$ as a stochastic dynamics, is
qualified in $\Sigma^\prime$ as strictly deterministic.
This phenomenon is depicted in figure \ref{Fig_11},
where  $v_h$ and $D_{h,k}$  are given by
$v_1=-0.5$, $v_2=0.433$, $D_{1,1}=0.208$, $D_{1,2}=0.060$, $D_{2,2}=0.0173$
(referred to as model II),
corresponding to the values of $v_h^\prime$ and $D_{h,k}^\prime$
obtained in the STD model depicted in figure
\ref{Fig_9} at $w=0.5$,
and $\Sigma^\prime$ moves with respect to $\Sigma$
with a negative relative velocity $w$.

\subsection{Observation 4 - Diffusivity and Markovian
processes in the Minkowski space-time}
\label{sec10_3bis}

The concept of tensor diffusivity for a relativistic
stochastic process is a long-term emerging property,
exactly as for Poisson-Kac processes  that are characterized
by an effective diffusivity for time-scales much larger than the
characteristic recombination time amongst its partial probability waves.

A local (pointwise) diffusivity (possessing the
dimension of m$^2$/s) cannot be defined in a Minkowski space-time
${\mathcal M}_4$ as it would be necessarily associated with
fluctuations possessing a local almost everywhere non-differentiable
structure as a function of time, and consequently an
unbounded local propagation velocity.

There is another general observation arising from
the analysis developed in Section \ref{sec9}.
related to the Markovian nature of a stochastic
process in a Minkowski space-time.

From the works by Dudley and Hakim quoted in the Introduction,
the impossibility of defining a strictly Markovian stochastic
kinematics (continuous in time) in ${\mathcal M}_4$ follows.

The case of STD processes introduced in Section \ref{sec9}
provides a concrete example of a random dynamics for which
the choice of a discrete time parametrization (the iteration
time $n$) makes it possible to define a strictly Markovian
process in ${\mathcal M}_4$, where space and time variables
are stricly treated on equal footing by defining the
space-time displacements $({\bf A}_h,\tau_h)$.
The example of STD processes does not contraddict the Dudley-Hakim
condition, as the extension with respect to a continuous
time variable of STD processes, associated with the
concept of hyperbolic homogenization \cite{giona_std2},
leads to non-strictly Markovian processes in $({\bf x},t)$
analogous to Poisson-Kac and Generalized Poisson-Kac processes.

\subsection{Observation 5 - Skewed structure of the transformation}
\label{sec10_4}

Given a stochastic process characterized by bounded propagation
velocity less than $c$, let ${\bf v}$ and ${\bf D}$ be its
effective (long-term) transport properties in a reference
frame $\Sigma$, and ${\bf v}^\prime$ and ${\bf D}^\prime$
the corresponding quantities measured in $\Sigma^\prime$,
moving with respect to $\Sigma$ with constant velocity $w$
along the $x_1$-axis. Here ${\bf D}$ and ${\bf D}^\prime$
are the diffusivity tensors in the two reference frames.

As regards the effective velocity, the transformation
from ${\bf v}$ to ${\bf v}^\prime$
is the classical velocity transformation of special
relativity
\begin{equation}
{\bf v}^\prime = {\mathcal V}_w[{\bf v}]
\label{eq10_8}
\end{equation}
for a relative velocity $w$. For the tensor
diffusivity the transformation expressed by
eqs. (\ref{eq9_28})-(\ref{eq9_29}) can be compactly 
indicated as
\begin{equation}
{\bf D}^\prime = {\mathcal D}_w[{\bf D},{\bf v}]
\label{eq10_9}
\end{equation}
Observe the skew-product structure of eq. (\ref{eq10_9})
in which the transformation for the diffusivity tensors
depends on ${\bf v}$. In a more compact form, eqs. 
(\ref{eq10_8})-(\ref{eq10_9}) can be summarized by the
complete transformation of the effective transport parameters
$({\bf v},{\bf D})$,
\begin{equation}
({\bf v}^\prime,{\bf D}^\prime)={\mathcal T}_w[({\bf v},{\bf D})]
\label{eq10_10}
\end{equation}
Obviously,
\begin{equation}
{\mathcal T}_0= \mbox{id} \, , \qquad {\mathcal T}_w^{-1}=
{\mathcal T}_{-w}
\label{eq10_11}
\end{equation}
Furthermore, the skew product nature of eq. (\ref{eq10_10})
implies
\begin{equation}
{\mathcal D}_{-w}[{\mathcal D}_w[{\bf D},{\bf v}],{\mathcal V}_{w}[{\bf v}]]
= {\bf D} \qquad \forall \,  {\bf v} \in (-c,c)^3
\label{eq10_12}
\end{equation}

\section{Concluding remarks}
\label{sec11}

The purely kinematic investigation of stochastic processes
in the Minkowski space-time opens up
interesting perspectives in the analysis of the
relativistic implications of stochasticity and determinism.
Using Poisson-Kac processes first, and STD dynamics subsequently,
the relativistic transformation of the tensor diffusivity
has been derived.

Particularly interesting are the conceptual implications
on the meaning of stochasticity in a relativistic framework,
which in some sense is frame dependent, i.e., it 
depends on the observer's velocity.

Two main observations should be pinpointed. The first is
the relativistic invariance of a quantity having the dimension
of an action, for a particle of rest mass $m_0$ moving
of purely ``diffusive'' motion, i.e.,
in the case where the effective convective contributions
are vanishing. Further analysis will clarify whether this
is just a nice coincidence or it admits more fundamental
quantum mechanical implications associated with the definition
of the Planck constant $h$.
The second observation is the fading of diffusivities
measured in inertial systems $\Sigma^\prime$ moving
with a relative velocity  $w$
approaching that of light, i.e., $\lim_{w \rightarrow c}
D_{h,k}^\prime(w)=0$. For an observer in $\Sigma^\prime$ all the
effects associated with stochasticity (e.g. irreversibility, dissipation,
etc.) are suppressed for $w \rightarrow c$. In this
framework, the concept of light velocity seems to
acquire a new thermodynamic meaning as  the threshold
velocity at which external  stochastic irreversible processes loose
their dissipative nature and approach a strictly deterministic
dynamics.
The extension of this, purely kinematic analysis of stochastic
processes
within the Riemannian space-time of general relativity will be
considered in a forthcoming contribution.

\end{document}